\documentclass[a4paper]{jpconf}
\usepackage{aas_macros}

\usepackage{natbib}
\usepackage[percent]{overpic}
\usepackage{graphicx}
\usepackage{epstopdf}
\usepackage{epsfig}
\usepackage{float}
\usepackage{fancyvrb} 
\usepackage{amssymb}

\usepackage[hyphens]{url}
\usepackage{hyperref}

\usepackage{tabularx}
\usepackage{tabulary}
\usepackage{array}
\usepackage{verbatim}

\usepackage{calc, layouts}
\usepackage{wrapfig}
\usepackage{citesort}

\begin{document}

\SaveVerb{WID}|WID|
\SaveVerb{WID=4}|WID=4|
\SaveVerb{WID=8}|WID=8|
\SaveVerb{rocprof}|rocprof|
\SaveVerb{WID3}|WID3|
\SaveVerb{globalToLocalMap}|globalToLocalMap|
\SaveVerb{SpatialCell}|SpatialCell|

\title[Vlasiator GPU portability]{Porting the grid-based 3D+3V hybrid-Vlasov kinetic plasma simulation Vlasiator to heterogeneous GPU architectures}

\author{Markus~Battarbee\textsuperscript{1}, 
Konstantinos~Papadakis\textsuperscript{1}, 
Urs~Ganse\textsuperscript{1}, \\
Jaro~Hokkanen\textsuperscript{2}, 
%Samuel~Antao\textsuperscript{3}, 
Leo~Kotipalo\textsuperscript{1}, 
Yann~Pfau-Kempf\textsuperscript{1}, 
Markku~Alho\textsuperscript{1}, 
and Minna~Palmroth\textsuperscript{1,4}
}
\address{$^1$ University of Helsinki, Department of Physics, P.O. Box 4, 00014 University of Helsinki, Finland 
\\
$^2$ CSC – IT CENTER FOR SCIENCE LTD. P.O. Box 405 FI-02101 Espoo, Finland
\\
%$^3$ Advanced Micro Devices, Inc.
%\\
$^4$ Finnish Meteorological Institute, Space and Earth Observation Centre, PO Box 503, 00101 Helsinki, Finland
}

\ead{markus.battarbee@helsinki.fi}
%\date{\today}

\begin{abstract}
Vlasiator is a space plasma simulation code which models near-Earth ion-kinetic dynamics in three spatial and three velocity dimensions. It is highly parallelized, modeling the Vlasov equation directly through the distribution function, discretized on a Cartesian grid, instead of the more common particle-in-cell approach. Modeling near-Earth space, plasma properties span several orders of magnitude in temperature, density, and magnetic field strength. In order to fit the required six-dimensional grids in memory, Vlasiator utilizes a sparse block-based velocity mesh, where chunks of velocity space are added or deleted based on the advection requirements of the Vlasov solver. In addition, the spatial mesh is adaptively refined through cell-based octree refinement. In this paper, we describe the design choices of porting Vlasiator to heterogeneous CPU/GPU architectures. We detail the memory management, algorithmic changes, and kernel construction as well as our unified codebase approach, resulting in portability to both NVIDIA and AMD hardware (CUDA and HIP languages, respectively). In particular, we showcase a highly parallel block adjustment approach allowing efficient re-ordering of a sparse velocity mesh. We detail pitfalls we have overcome and lay out a plan for optimization to facilitate future exascale simulations using multi-node GPU supercomputing.
\end{abstract}

%\maketitle

\section{Introduction}
\label{sec:introduction}

Vlasiator is a hybrid-Vlasov space plasma simulation code, specifically designed to model kinetic plasma dynamics of the near-Earth environment \citep{Palmroth_2018_LRCA,Ganse_2023_POP6D}. Vlasiator has been run on some of the largest supercomputers in the world, as due to the high fidelity of its model, simulations are computationally costly, typically exceeding millions of CPU-hours for a few tens of minutes of simulated physical time. Vlasiator, in its CPU iteration, utilizes vector instructions (AVX2, AVX512), OpenMP threading, as well as the message passing interface \citep[MPI,][]{walker1996mpi} to run on distributed hardware, supporting several hundreds of nodes and tens of thousands of MPI tasks. The field of supercomputing has seen a transition from nodes utilizing standard central processing units (CPUs) to ones including a variety of general purpose graphics processing units (GPGPUs, or GPUs for short) \citep{asaduzzaman2021impact}  for acceleration of mathematical operations and access to high-bandwidth memory \citep[HBM,][]{8293707}.

Porting a supercomputer simulation code to heterogeneous GPU/CPU hardware has a two-fold goal. A first objective is to be able to run simulations on a hardware previously not supported by the codebase, yet offering tempting performance potential. A second, yet in many ways more important objective is to refactor and optimize the code so the new hardware can be exploited to its fullest, facilitating faster, longer, or more detailed simulations than would be possible with previous-generation CPU hardware. Altering the simulation code to support GPU hardware can be achieved through various approaches discussed later in this manuscript. In this paper, we focus on two approaches, which are the Compute Unified Device Architecture \citep[CUDA,][]{luebke2008cuda} by NVIDIA corporation, and the C++ Heterogeneous-Compute Interface for Portability (HIP) by Advanced Micro Devices, Inc.\ (AMD) which acts as an interface to the lower-level ROCm software stack.

In this paper, we describe the architectural choices and resultant changes implemented in the Vlasiator simulation code during its first stages of porting to support GPU architectures via the CUDA and HIP/ROCm interfaces.

\subsection{Vlasiator}\label{sec:vlasiator}

% - Code overview and introduction
 Hybrid ion-kinetic plasma simulations such as Vlasiator model the evolution of ions, assuming electrons are a massless charge-neutralizing fluid. Vlasiator simulates collisionless plasma physics through evolving the ion (usually proton) velocity distribution function (VDF) directly instead of relying on macroparticle sampling as used by particle-in-cell \citep[PIC,][]{nishikawa_pic_2021}) approaches. The evolution of the VDF is evaluated through solving the collisionless Boltzmann equation, also known as Vlasov equation \citep{Vlasov_1961_Vlasov},
\begin{equation}
  \frac{\partial f_s}{\partial \mathrm{t}} + \mathrm{\vec{v}} \cdot \frac{\partial f_s}{\partial \mathrm{\vec{x}}} + \frac{\mathrm{q}_s}{\mathrm{m}_s}
  \left( \vec{E} + \mathrm{\vec{v}} \times \vec{B} \right) \cdot
  \frac{\partial f_s}{\partial \mathrm{\vec{v}}}
      = 0.
      \label{eq:vlasov}
\end{equation}
where $f_s\left(\vec{\mathrm{x}},\vec{\mathrm{v}},t\right)$ is the phase-space density for ion species $s$ (with charge $\mathrm{q}_s$ and mass $\mathrm{m}_s$), and $\vec{E}$ and $\vec{B}$ are the electric and magnetic fields, respectively. Using a Vlasov approach has many benefits such as a representation of VDFs free of sampling noise and the capability to simulate correct scale separation between the magnetic obstacle of the Earth and ion-kinetic effects, but is numerically expensive to compute. Vlasiator simulates the global magnetospheric environment of Earth with the simulation domain containing inflowing solar wind up to a distance of several tens of Earth radii \mbox{($1\,R_\mathrm{E} \approx 6371\,\mathrm{km}$)}, the bow shock and foreshock regions, the magnetosheath and the cusps, as well as the flanks and the magnetotail to a distance of several tens of $R_\mathrm{E}$.

Vlasiator utilizes explicit solvers with a Strang splitting \citep{Strang_1968_Splitting,Einkemmer_2014_StrangSplitting} approach, decomposing evolution of (\ref{eq:vlasov}) into six Cartesian shear operations; three in velocity space (referred to as \emph{acceleration}), and three in spatial space (referred to as \emph{translation}). The Lorentz force effect on the plasma distribution (first and third terms of equation \ref{eq:vlasov}) can be (in the non-relativistic case) described as gyration in a frame of reference where the motional electric field cancels out the existing electric field. Using the SLICE-3D \citep{Zerroukat_2012_SLICE3D} algorithm, this is deconstructed into three Cartesian direction aligned shear transformations. 

\subsection{Discretization}

Vlasiator discretizes ion VDFs on Cartesian grids, storing and propagating phase-space average values in units of \mbox{$(\Delta x \Delta y \Delta z \Delta v_x \Delta v_y \Delta v_z)^{-1}$}. The simulation domain consists of a 2D or 3D spatial Cartesian mesh, which can be either uniform, or use cell-based mesh refinement with an octree approach. For 2D simulations, the third dimension consists usually of a thickness of a single cell with periodic boundary conditions. Non-unitary thicknesses (so-called 2.9D simulations with a cylindrical inner boundary) are also supported, such as those used for a study of dayside reconnection with multiple X-lines \citep{PfauKempf_2020_2.9D}. When simulating in three spatial dimensions, the octree approach is used to refine low-resolution cubic cells into 8 children each with half the dimensions of the parent. Within a neighbourhood of 3 cells in each direction, refinement level can change by only one level, to prevent too drastic resolution boundaries. Refinement can be repeated several times, with current simulations utilizing up to 4 different resolution levels. Details of the AMR approach used in Vlasiator are presented in \citet{Papadakis_2022}, \citet{Ganse_2023_POP6D} and \citet{Kotipalo_2024_dAMR}. Typical Vlasiator 2D (or 5D; 2D+3V=5D) simulation spatial resolutions have been between $227$ and $300\,\mathrm{km}$, slightly above the ion inertial length in the solar wind, but capable of modelling ion-kinetic effects as shown in \citet{PfauKempf_2018_scaling}. In 6D runs, the highest spatial resolution in regions of interest has been $1000\,\mathrm{km}$ which is above the ion inertial length, but those simulations have been shown to still include significant ion-kinetic dynamics \citep{Grandin2023,Palmroth2023}.

The electric and magnetic fields are solved on a Cartesian uniform Yee-lattice \citep{yee_numerical_1966} grid, with resolution matching that of the highest refinement level of the Vlasov grid. This approach ensures the field solver encounters no resolution boundaries, which could potentially lead to reflection or refraction of propagating electromagnetic wave modes \citep{FrankReich2004}. Fields are propagated using Maxwell-Amp\`ere's law
\begin{equation}
  \nabla\times\mathbf{B} = \mu_0\mathbf{j} + \frac{1}{c^2}\frac{\partial \mathbf{E}}{\partial t},
\label{eq:Ampere}
\end{equation}
and Faraday's law
\begin{equation}
  \nabla\times\mathbf{E} = -\frac{\partial \mathbf{B}}{\partial t},
\label{eq:Faraday}
\end{equation}
under the Darwin approzimation (in effect, neglecting the displacement current which is the last term of Eq.~\ref{eq:Ampere}), with closure provided by Ohm's law
\begin{equation}
  \mathbf{E} + \mathbf{V}\times\mathbf{B} = \frac{\mathbf{j}}{\sigma} + \frac{\mathbf{j}\times\mathbf{B}}{n_\mathrm{e} e} - \frac{\nabla\cdot\mathcal{P}_\mathrm{e}}{n_\mathrm{e} e} + \frac{m_\mathrm{e}}{n_\mathrm{e} e^2}\frac{\partial \mathbf{j}}{\partial t},
\label{eq:Ohm}
\end{equation}
with the Hall and electron pressure gradient terms included, and assuming infinite conductivity ($\sigma \rightarrow \infty$) as well as adiabatic electrons ($\mathcal{P}_\mathrm{e} \propto n_\mathrm{e}^{5/3} $) with no inertia ($m_\mathrm{e} \rightarrow 0$); neglecting the first and last terms on the right-hand side. The discretized solver approach described in \citet{Londrillo_2004_fieldsolver} conserves magnetic field divergence to machine precision, thus obviating the need for any divergence cleaning step for the magnetic field $\vec{B}$ provided initial conditions are divergence-free.

\begin{figure}[htp]
    \includegraphics[width=0.8\textwidth]{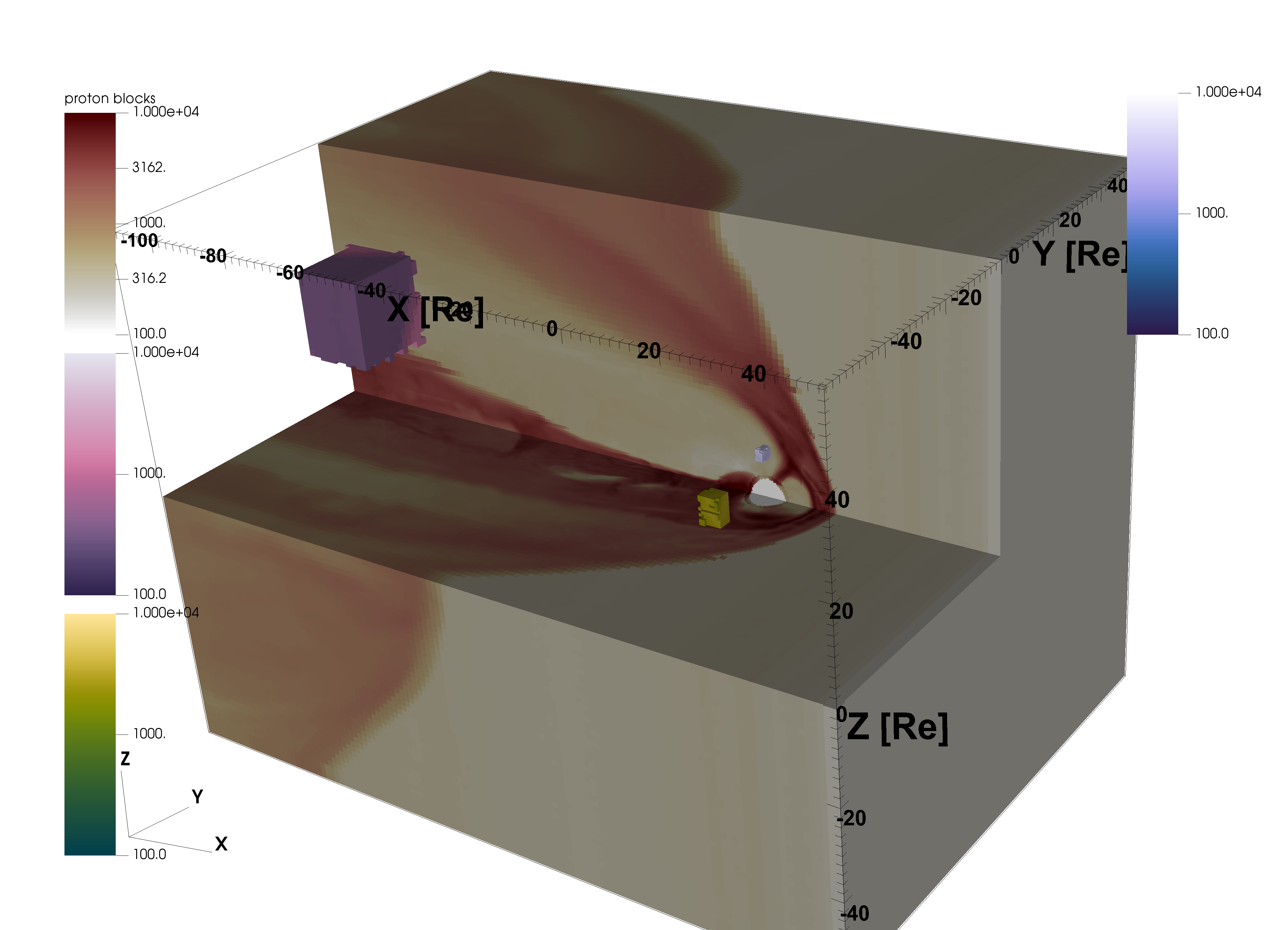}
    \caption{Example of grid decomposition for a global magnetospheric simulation. The figure shows three quadrants of simulation data with the reddish colour map indicating the local count of blocks. Each block contains \protect\UseVerb{WID3} (in this case 64) phase-space cells, and thus, the block count is a measure of velocity space complexity in each spatial position. In addition, the spatial DCCRG domains of three specific MPI tasks are shown: In purple, one domain in the corner of the simulation covers a large region with cells at low refinement levels. Within the magnetosheath, an intermediate-size region in greenish yellow contains non-maximally refined spatial cells with elevated block counts. Finally, the very small blue region (north of the inner boundary) consists of maximally refined spatial cells with very large block counts indicating a high computational load per cell.}
    \label{fig:DCCRG_example}
\end{figure}

Each spatial cell on the spatial mesh contains a local representation of phase-space, also discretized on a Cartesian velocity mesh. To conserve memory and computational resources, Vlasiator utilizes a sparse velocity space approach, only storing and propagating those regions of velocity space which contain non-negligible phase-space densities. This is performed by declaring the maximum allowable velocity space extent and dividing it into cubic \emph{blocks} of \mbox{$(4 \times 4 \times 4)$} or, as implemented recently, \mbox{$(8 \times 8 \times 8)$} velocity space cells each. Internally this is declared with notation of \verb|WID=4| or \verb|WID=8|, with values of \Verb|WID3=64| and \Verb|WID3=512|, respectively. Typical maximal velocity space extents have been $\pm 4.02\cdot10^6\,\mathrm{m s}^{-1}$ in each direction with 67 blocks per direction, resulting in a velocity space resolution of $30\,\mathrm{km s}^{-1}$, or with 6D simulations, $\pm 4\cdot10^6\,\mathrm{m s}^{-1}$ with 50 blocks per direction, leading to a velocity space resolution of $40\,\mathrm{km s}^{-1}$. This reduction of velocity resolution hand-in-hand with spatial resolution follows the findings of \cite{PfauKempf_2018_scaling}. Active blocks are added to the velocity space or removed from it as necessary, as described in \citet{Alfthan_2014_vlasiator}. We ephasize that the term sparse is used in this manuscript to discuss this mechanism for partitioning and selectively storing and propagating portions of potential velocity space, and thus it is not directly related to e.g.\ sparse matrix algorithms. An example of a Vlasiator simulation domain, showcasing spatial mesh refinement and differing spatial cell computational weights due to varying block counts, is shown in Figure \ref{fig:DCCRG_example}. 

%Vlasiator has been running on top-end CPU-based supercomputers for over 10 years, but with pre-exascale and early exascale supercomputers such as LUMI and FRONTIER relying on GPUs for the bulk of their computational capacity, Vlasiator is also being rewritten to support GPU accelerator hardware. This manuscript details the architectural choices and results of this process.

\section{Vlasiator data structures}\label{sec:datastructures}

\begin{figure}[htp]
    \includegraphics[width=0.6\textwidth]{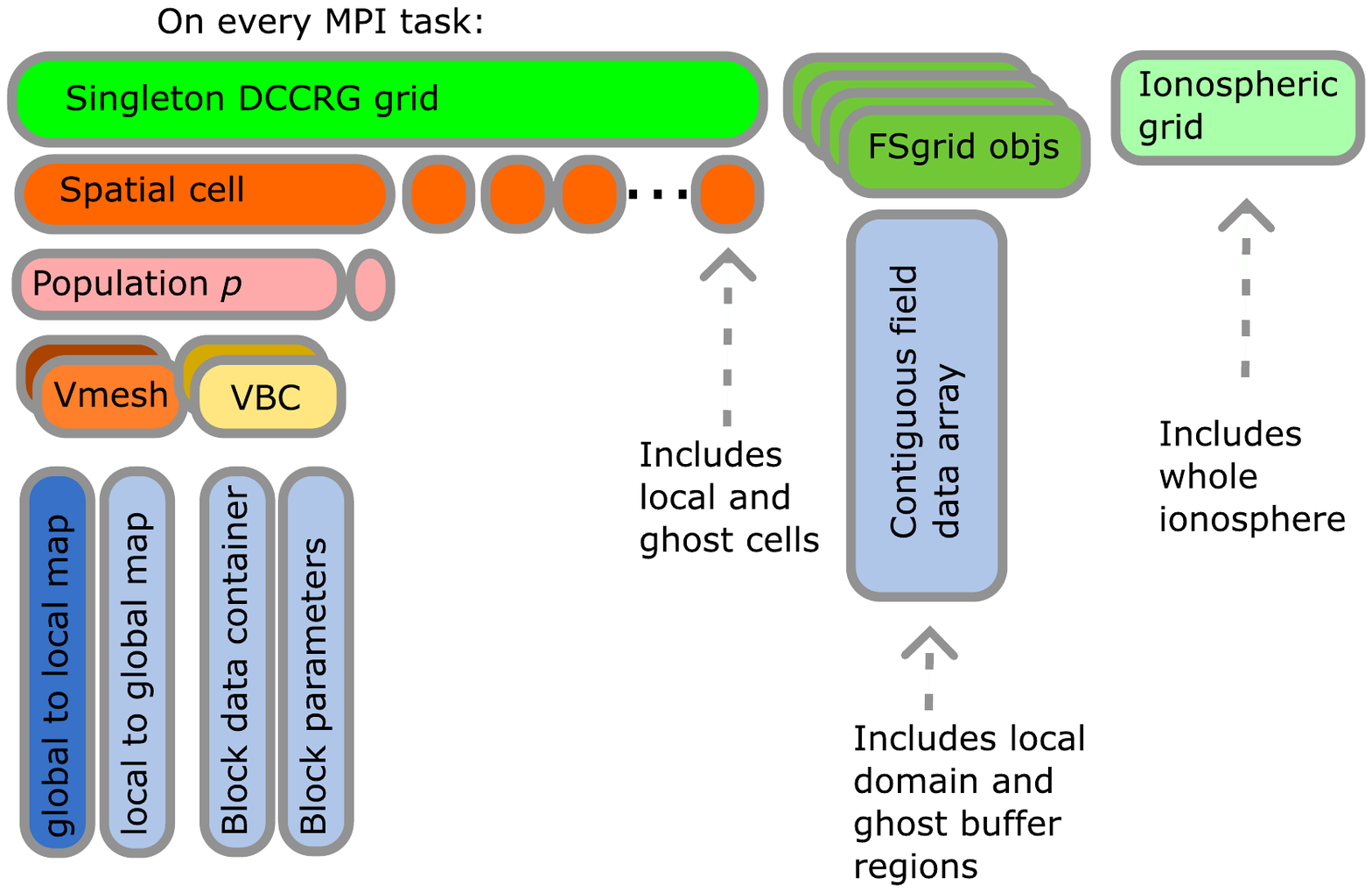}
    \caption{Schematic of data constructs within each Vlasiator MPI task. Each task contains a DCCRG grid object, which holds a variable number of \protect\UseVerb{SpatialCell} objects. Each spatial cell will contain one or more particle populations (declared on simulation setup), and each population in turn contains a Velocity Mesh (Vmesh) and a Velocity Block Container (VBC). The Vmesh links memory buffer positions (LIDs) with physical velocity coordinates (GIDs). The \protect\UseVerb{globalToLocalMap} is a hash map (dark blue), with GIDs as keys and LIDs as values. The complementary data structure is a vector (light blue) with size equal to that of the map, and at each index (LID) it stores the GID which corresponds to that index. The VBC, on the other hand, contains the actual VDF data (Block data container, vector, light blue) as \protect\UseVerb{WID3} floating point values for each LID, as well as pre-calculated metadata (Block parameters, vector, light blue) for the \protect\UseVerb{WID3}-sized block in question. On GPU architectures, Vmesh and VBC wrapper objects are duplicated on both host and device memories. Cartesian FSgrid objects each contain a contiguous, fixed-length vector data buffer, declared in host memory, for use by the field solver. Additionally, each MPI task contains a whole ionospheric grid for handling the inner boundary.}
    \label{fig:Data_constructs}
\end{figure}

Vlasiator utilizes three different in-house developed grid libraries. The data objects associated with each grid are presented in Figure \ref{fig:Data_constructs}. The first, the Distributed Cartesian Cell-Refinable Grid \citep[DCCRG,][]{Honkonen_2013_DCCRG}, houses the particle phase-space data on an octree-cell-refinable spatial mesh. The mesh discretizes the spatial domain in up to three dimensions with each spatial cell containing VDFs for all velocities at a given spatial location. Each spatial cell is represented by a C++ class \Verb|SpatialCell| object as well as additional metadata such as information on the cell resolution and neighbourhood. The DCCRG grid is distributed among MPI tasks based on cell weights, where each cell has its weight determined as a function of how complex a velocity space it contains. This load distribution is performed by the Zoltan library \citep{Boman_2012_Zoltan}, which supports several different domain decomposition algorithms such as Recursive Coordinate Bisection (RCB) or Recursive Inertial Bisection (RIB). The grid object of each MPI task contains all spatial cells which have been assigned to the task in question, as well as copies of any neighbouring cells residing on remote tasks, and implements an interface for updating local copies of remote (ghost) cells. Due to the use of Strang splitting, the chosen MPI decomposition ensures that all velocity-space acceleration operations (using the first and third terms on the left-hand side of Equation \ref{eq:vlasov}) use only local information. This also implies that any adjustments to the local sparse velocity space grid are performed using local data. An exception to this rule is that cells will also need to ensure they have buffer data so that during spatial advection (solving the first and second terms on the left-hand side of Equation \ref{eq:vlasov}) phase-space density $f_s$ has a target to flow into.

Each \Verb|SpatialCell| object contains, among other things, local average information such as plasma moments and largest locally allowed time propagation steps, as well as a vector of particle populations. Vlasiator is usually run with one species of ions, namely protons, and with electrons providing a charge-neutralizing fluid. However, multiple ion populations such as helium as described in \cite{Battarbee_2020_Helium} are also supported. The specialized eVlasiator model \citep{Battarbee_2021_evlasiator} also maintains two particle populations (electrons and ions) but populates only the electrons. Each of the particle population (declared at the start of the simulation) keeps track of population-specific velocity moments (density, bulk velocity, pressure tensor) and contains descriptors for the VDF contained in that spatial location. As Vlasiator propagates phase-space density $f_s$, the value contained in each spatial cell is not directly impacted by the cell size, but rather, always represents the spatial average over the local cell volume. 

\subsection{Sparse block representation in memory}

The VDF at a given spatial location is represented by two C++ classes: a \emph{Velocity mesh} (Vmesh) and a \emph{Velocity Block Container} (VBC). The Vmesh links memory buffer positions (local address offset i.e.\ the local block identifier, LID) with physical velocity coordinates (global block identifier, GID). It contains a hash map (\emph{globalToLocalMap}) with GIDs as keys and LIDs as values. The complementary data structure is a vector (\emph{localToGlobalMap}) with size equal to that of the map, and where each index (LID) stores the corresponding GID. The VBC, on the other hand, contains the actual VDF phase-space data $f_s$ (Block data container, vector) as \Verb|WID3| floating point phase-space density values for each block and LID, as well as pre-calculated metadata (Block parameters, vector) for each \Verb|WID3|-sized block in question. LIDs are thus simply indices into \Verb|WID3|-staggered memory space, whereas GIDs represent a position in velocity space, with values starting from 0 for the smallest velocity value in all Cartesian directions, incrementing first along the $V_x$ axis, then along the $V_y$ axis, and finally along the $V_z$ axis. When evaluating 6-dimensional phase-space neighbours of any given block, spatial neighbours will have the same GID, whereas velocity neighbours will have a GID differing by $\pm 1$, $\pm N_{\mathrm{b},x}$, or $\pm N_{\mathrm{b},x} N_{\mathrm{b},y}$ where $N_{\mathrm{b},i}$ is the maximum possible number of blocks in Cartesian direction $\mathrm{i}$.

\subsection{Block adjustment requirements}

\begin{figure}[htp]
    \includegraphics[width=0.7\textwidth]{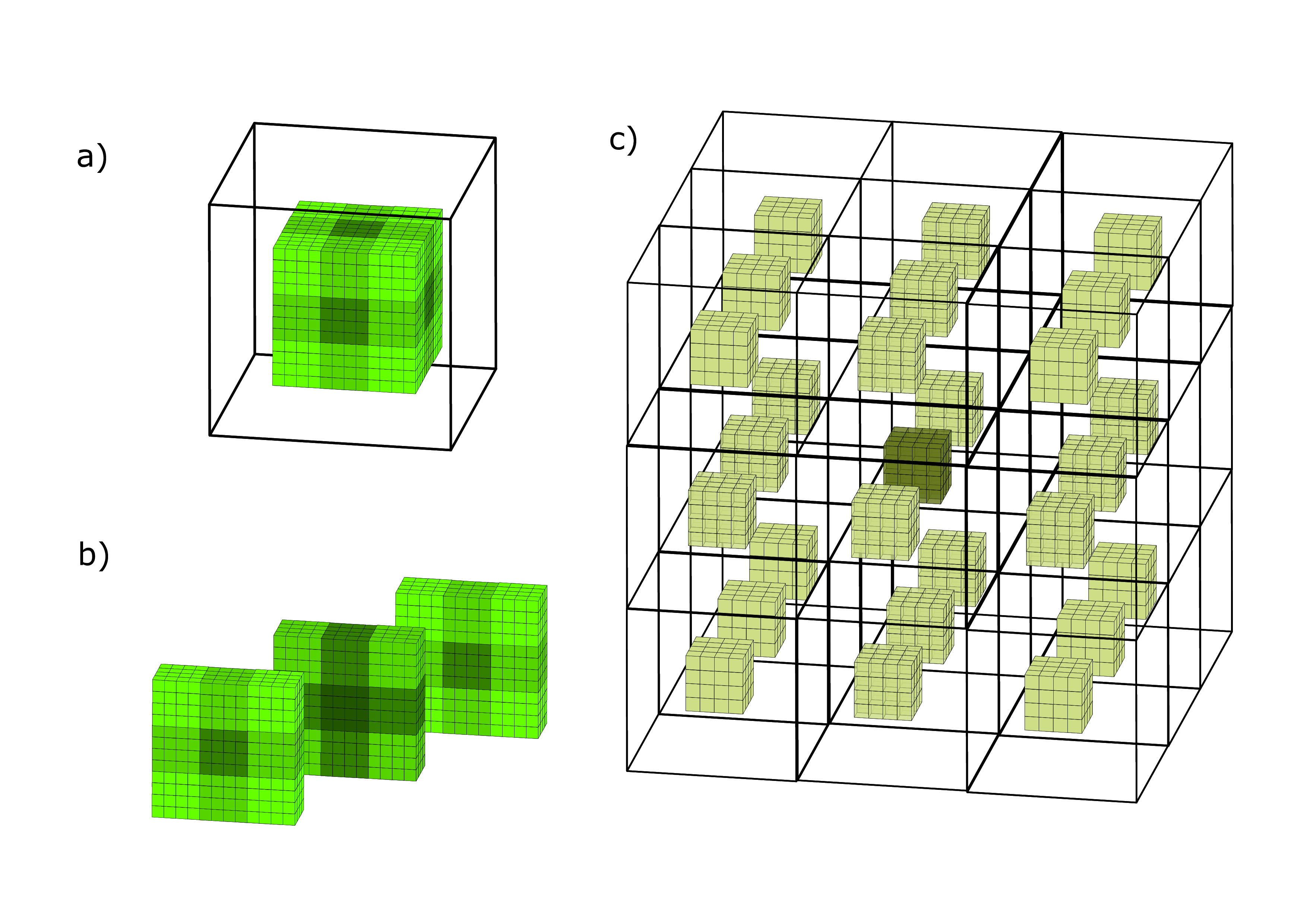}
    \caption{Ghost requirements of the Vlasiator Vlasov solvers. Each $(4 \times 4 \times 4)$ velocity block which has phase-space density above the sparsity threshold $f_\mathrm{th}$ (center block in dark green) requires neighbours for plasma to flow into. 
a) Neighbours in velocity space can be face neighbours (medium green, 6 blocks) or edge or corner neighbours (paler green colours, 20 blocks). 
b) Deconstruction of panel a.
c) Each block above $f_\mathrm{th}$ requires existing neighbours also in spatial space. This panel demonstrates 26 spatial neighbour cells (tan) from an uniform 3D-3V simulation which all must contain the required velocity block, even if the central block (olive) were the only block with non-negligible phase-space densities. At spatial mesh refinement interfaces, the number of spatial neighbours may be larger than 26 (neighbours with higher refinement levels) or smaller than 26 (neighbours with smaller refinement levels).
    }
    \label{fig:Block_Neighbours}
\end{figure}

A naïve estimation of memory consumption for a 6D Vlasiator simulation with fully populated phase-space grids is $3.5 \times 10^{15}$ phase space cells, a memory requirement of about 14 petabytes \citep{Ganse_2023_POP6D,Kotipalo_2024_dAMR}. A significant contributor to Vlasiator's success is the sparse velocity mesh, reducing the computational domain to include only those regions of phase-space where non-negligible density $f_s > f_\mathrm{th}$ is found, yielding a memory reduction of over two orders of magnitude. 
This strength comes with the cost of having to constantly adjust and update the structure of phase-space stored in each cell. The acceleration solver evaluates future velocity space extents after performing each shear operation, flagging content blocks which are no longer needed and those which need to be created. The translation solver on the other hand requires that neighbouring spatial cells contain the required data structures for phase-space density to flow into. These requirements are described as the ghost halo regions, as indicated in Figure \ref{fig:Block_Neighbours}. After running the acceleration solver, each content block is flagged as being either above the sparsity threshold $f_\mathrm{th}$ (i.e.\ having significant content), or below it (i.e.\ not having significant content). Then, the halo requirements are evaluated for all content blocks, both for velocity and spatial neighbours. Finally, required halo blocks are created, and non-content blocks not required by any halos are deleted. 

\begin{figure*}
   \centering
   \includegraphics[width=0.8\textwidth]{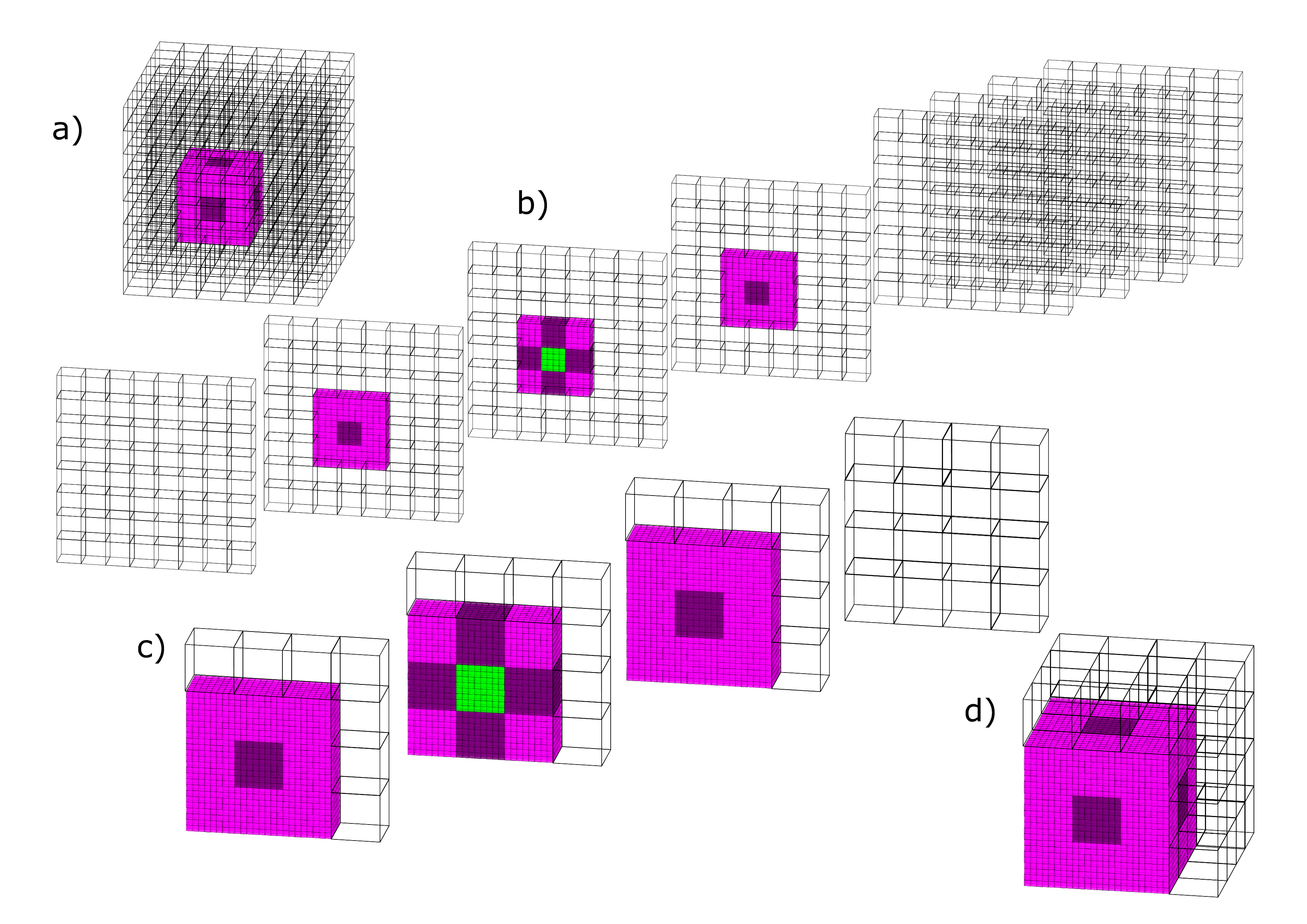}
   \caption{Illustration of the block-based velocity mesh with fixed total extents but a varying value for \protect\UseVerb{WID}. Panel a): Velocity domain consisting of blocks of $(4 \times 4 \times 4)$ cells. A green block (hidden) is flagged as containing significant plasma density, here assumed to be a point-like distribution. This causes neighbouring blocks (face neighbours in dark purple, edge and corner neighbours in fuchsia) to be retained in the sparse mesh structure. b): Exploded view of the velocity mesh, revealing the green content block. c): Exploded view of a mesh consisting of $(8 \times 8 \times 8)$ cell blocks instead of $(4 \times 4 \times 4)$ cell blocks, conscructed in response to the same point-like distribution as in panels a) and b). d) Velocity domain consisting of blocks of $(8 \times 8 \times 8)$ cells. Increasing block size reduces the amount of metadata required for representing the velocity grid, but on average increases the amount of velocity cells used to represent a given velocity distribution function.}
   \label{fig:wid4to8}
\end{figure*}

Managing velocity space sparsity via blocks allows Vlasiator to use standard data units of size \Verb|WID3|, useful for vectorization and GPU approaches. Vlasiator has now been amended to support blocks of size $(8 \times 8 \times 8)$ (\Verb|WID3|=512) in addition to the longstanding $(4 \times 4 \times 4)$  (\Verb|WID3|=64) block size. This results in much less overhead from creating and deleting blocks since a smaller number of blocks is required to describe any given phase-space distribution function, assuming the resolution remains the same. At the same time, the halo requirements will, on average, lead to a larger memory footprint and more actual phase-space cells to be propagated. This is a trade-off, but one which is expected to prove beneficial on GPU architectures. The implications on halos of changing the value of \Verb|WID| for a simple configuration, whilst maintaining constant maximum velocity space extents, is exemplified in Figure \ref{fig:wid4to8}.

\subsection{Other grid objects in memory}

In addition to the DCCRG grid containing spatial cells with particle species VDF data, each MPI task also owns several FSgrid objects which house uniform Cartesian rectangular domains of scalars or vectors for field solver use. The total spatial domain is decomposed uniformly over all MPI tasks, with addition ghost buffer regions around local data, which can be updated from the up to 26 neighbouring tasks. The division into several different FSgrid objects is to facilitate ghost communication of only the necessary data for each step of the field solver operation. The Vlasiator field solver, consisting of much smaller data domains than the Vlasov solver, has yet to be ported to heterogeneous GPU architectures.

A final data object of Vlasiator is the ionospheric grid \citep{ganse2024ionosphere}, which is contained on every task in its entirety and does not contain any MPI interface. The ionospheric grid exists only when simulating a magnetospheric run with an ionospheric inner boundary, and its solvers have yet to be ported to GPU architectures.

\subsection{Heterogeneous memory decomposition of data structures} \label{sec:memory_api}

As Figure \ref{fig:Data_constructs} introduced in Section \ref{sec:datastructures} shows, the data structures of Vlasiator are non-trivial when considering heterogeneous memory architectures such as those encountered in modern GPU-powered supercomputing nodes \citep{7446077,10.1145/2591971.2592002}. Computational activities performed by the CPU can usually only access host-side memory, and parallel computation kernels utilizing GPU hardware can only efficiently access device-side memory. Current pre-exascale and exascale supercomputers are built upon node architectures which house discrete GPUs controlled by standard CPUs, each with discrete memory domains. 
%Under CUDA architectures, the host and device memory address spaces are discrete, whereas in AMD hardware the memory space is continuous. 
GPUs are built using high-bandwidth memory  \citep[HBM,][]{8293707}, allowing rapid availability of data for GPU computations with speeds of up to 3.2 TB/s on MI250x and 2 TB/S on NVIDIA A100. A bottleneck, however, is the connection between GPU and CPU hardware, currently limited to 
%\mbox{PCI Express 4.0 or 5.0} speeds of up to \mbox{64 or 128 GB/s}, 
speeds of tens to a few hundreds of GB/s per direction (depending on NVLink or PCI Express implementation), 
necessitating planned and controlled movement of computation-critical data between the two memory subsystems. Upcoming hardware such as the Nvidia Grace Hopper GH200 will upgrade this to a 900 GB/s bidirectional NVLink-C2C connection as well as a transparent unified memory domain doing away with the need for data migration. GPU HBM bandwidth will still remain superior at 3.0 TB/s. Accelerated processing unit (APU) chips such as the AMD Instinct MI300A will contain a single memory domain \citep{7920816}, which may simplify memory management on future exascale architectures whilst simultaneously providing massive bandwidth (5.3 TB/s on the MI300A).
% NV A100 and AMD MI250x both  	PCIe 4.0 x16

GPU application programming interfaces (APIs) provide access methods allowing various ways of copying data between host and device memory regions. A user-friendly implementation (supported on modern architectures such as NVIDIA Pascal and later and AMD MI200+, which still have physically distinct memory domains) is that of Managed Memory or Unified Memory (UM), where a single virtual memory address can point, as necessary, to either host or device memory. The UM subsystem will utilize a paging mechanism, and can even allow oversubscription of device memory, paging (migrating) memory to and from the device as necessary. Whenever the CPU or GPU needs to access data residing in UM, the paging system will be queried for the locality of the data. If the data is not in the required region, a \emph{page fault} occurs, triggering migration of the data to the correct domain. To reduce page faulting (which can harm performance), users can \emph{prefetch} data manually to the desired domains. One noteworthy aspect of prefetching and page faulting is that the UM system only migrates data in units of the page size, which can lead to unintended migration of data units residing within the same page as an intentionally migrated data unit. However, the benefit of unified memory is that an unit of data which is accessed from both host and device is ensured to always contain the correct value, irrespective of whether the user has declared prefetches. Should a data construct contain separate host-only and device-only memory buffers, it is the responsibility of the user to perform manual memory copies in order to maintain correct values for host and device activities, respectively. Additionally, automatic oversubscription of device memory is not possible if separate buffers are used. When porting a large simulation software such as Vlasiator, this reliability of UM was deemed a high priority, and thus, many Vlasiator data structures were migrated to UM instead of maintaining pure host and device copies of data buffers.

Of the data containers described above and showcased in Figure \ref{fig:Data_constructs}, only the lowest level of data containers, i.e.\ the hashmaps and vectors (dark and light blue in the figure) reside in unified memory. This includes both the data buffers and also metadata such as the container sizes. For FSgrid data, both unified memory and dual host-and-device memory versions have been implemented. To facilitate access to Vlasov data structures within the Vmesh and VBC objects from both host and device environments, Vlasiator stores discrete host and device copies of the actual Vmesh and VBC objects, which both point to the same hash maps and vectors residing in UM.

\subsection{Data structure re-ordering for Vlasov solver use} \label{sec:re-ordering}

%The Semi-Lagrangian update scheme utilized in Vlasiator is described in detail in \citet{Palmroth_2018_LRCA} with specific application to mesh-refined spatial space described in \citet{Ganse_2023_POP6D}. In short, the approach 
The Vlasiator Vlasov solver approach, introduced in Section \ref{sec:vlasiator}, 
is to decompose the Vlasov equation into six one-dimensional Cartesian shear or advection operations. Each advection operation takes a number of sampling points in the relevant direction, using them to fit a polynomial which describes the variation of phase-space density $f_s$ along the direction of propagation. This polynomial is then advected with relevant speed and direction and integrated over the target phase-space cell dimensions.
%length. Each act of advection, performed on one cell, will result in some phase-space density $f'_s$ flowing to the positive direction neighbour, with the local cell retaining the difference $f_s - f'_s$. The updated cell value thus consists of the retained local value and the increment flowing in from a negative-direction neighbour.

%Spatial advection or \emph{translation} utilizes the appropriate velocity space coordinate for each phase-space cell, with neighbours existing in the spatial domain. This in effect performs evolution according to the first and second left-hand side terms in equation (\ref{eq:vlasov}). In velocity space, we transform the acceleration and gyration as described by the Lorentz force in the first and third left-hand side terms of equation (\ref{eq:vlasov}) into three Cartesian-aligned shear operations, according to the SLICE-3D method described in \citet{Zerroukat_2012_SLICE3D}. As Vlasiator is constrained to the non-relativistic regime, we perform a velocity transformation into a frame in which the local electric field vanishes, resulting in pure gyration around the magnetic field. The advection velocity for each phase-space cell is then constructed from the gyration axis direction, the gyration angle performed over one timestep, and the transformation matrix used to transform to the gyration frame.

For each advection thus performed by our semi-Lagrangian approach, we require a varying number of control points depending on the polynomial reconstruction order. In order to limit the number of ghost cells required, we usually constrain translation to use parabolic polynomials for each cell, whereas acceleration utilizes quadratic polynomials. This results in each advection calculation requiring both the local phase-space density $f_s$ value and values of a number of neighbours in the direction of advection. For this reason, the first step of each Vlasov solver cycle is to take the existing Vlasov data structure and 
re-arrange it so that memory access operations will occur on contiguous and squential data regions.
%sort it so that the blocks of our calculation unit with size \Verb|VECL| are re-ordered to be sequential with respect to the advection direction.

For spatial translation, the solver will act on one global block identifier (GID) at a time, collecting the data along the translated direction, utilizing the pencil formalism described in \citet{Ganse_2023_POP6D}. For each spatial cell, the GID being translated will be linked to the local block identifier (LID) with the \verb|globalToLocalMap| hashmap object inside the Vmesh. The block data from each memory address given by the LID is then copied into a temporary device-only buffer, providing the solver with aligned block data along the direction of propagation. 
%, in units of size \Verb|VECL|. 
The original data buffers pointed to by each LID are then emptied (written to zero), and the semi-Lagrangian propagator is then able to write its output into the original data locations.

For velocity-space acceleration, phase-space densities will be advected along one Cartesian velocity direction at a time. The GID associated with each data block is such that values increase in the order \mbox{$V_x \rightarrow V_y \rightarrow V_z$}. In order to ensure the block data is aligned along the direction of advection $i$, we compute a transposed $\widetilde{GID}_i$ where $\widetilde{GID}_x=GID$. $\widetilde{GID}_y$ is defined by the order \mbox{$V_y \rightarrow V_z \rightarrow V_x$} and $\widetilde{GID}_z$ by the order \mbox{$V_z \rightarrow V_x \rightarrow V_y$}. In order to build the ordered column data structures, we utilize a sorting algorithm on the $\widetilde{GID}$-values, thus gaining access to the correctly sorted LIDs as well. The data from these addresses is then copied to a temporary buffer, which can be fed to the Vlasov solver. In the meanwhile, the contents of the velocity blocks are reset, so that the Vlasov solver can write the results of the solver operation into the cell data contents.

\section{Porting to GPUs: assessing different approaches}

In this section we will describe the approaches considered for porting of Vlasiator to GPU-accelerated architectures along with lessons learned. The observant reader may note that the basic Vlasiator data structure applied to VDFs, being a block of size \Verb|WID3|, is beneficial for performing parallel calculations. Historically Vlasiator has utilized a value of \Verb|WID3|=64, which indicates that two Nvidia 32-element warps or a single AMD 64-element wavefront can perform all necessary evolution calculations for a whole block of data. This stems from an early finite volume method version of Vlasiator, which included a CUDA interface, as presented in \cite{SANDROOS2013306}. Sadly, when Vlasiator transitioned to the current Semi-Lagrangian solver \citep{Palmroth_2018_LRCA}, the old CUDA support was deprecated. The benefit of an internal block-based data structure which is easily applicable to SIMT approaches, however, has remained a hallmark of Vlasiator ever since, and has been utilized via single instruction multiple data (SIMD) vector operations on CPUs with great success.

\subsection{Directive-based approaches}

Vlasiator, utilizing existing CPU-based computational facilities to a great degree, is built to perform calculations both in parallel distributed over MPI to several tasks, as well as running several OpenMP threads in parallel, possibly also benefiting from hyperthreading. As such, OpenMP directives are implemented in several parts of the Vlasiator codebase, and directives were an interesting direction for trialling GPU porting. During the autumns of 2019 and 2020, the Vlasiator development team participated in Hackathons where OpenACC directive-based GPU programming was tutored and implemented. However, our testing suggested that OpenACC was not going to provide us with sufficient throughput, as mere pragma-based offloading of selected code sections would not easily result in sufficient re-factoring or provide us with the granularity of memory management foreseen to be required. Vlasiator, implementing a 6-dimensional advection equation, requires each point of VDF data to be available for each solver step, save for certain ghost cell information, meaning VDF data must remain on the device. Additionally, compiler support of pragma-based approaches during our initial porting efforts was lacking. As such, we did not pursue OpenACC approaches further, nor did we trial \mbox{OpenMP 5.0} GPU-based directives.

\subsection{Abstraction layers}
%Sidestep: Disregarding abstraction layers such as Kokkos, Alpaka, and RAJA

A useful approach in porting existing software to new architectures is to utilize a portable abstraction layer, which can be externally maintained and updated to allow interfacing with new hardware and programming languages without the end-user needing to change their code. Abstraction can be built into the grid library itself \citep[such as in AMReX,][]{AMReX_JOSS} or provided through an external portability layer, which in turn may depend on libraries for memory management. Examples of such portability layers include RAJA (\citealt{RAJA} with array abstractions via CHAI and memory management via UMPIRE \citet{umpire}), alpaka (\citealt{alpaka_MathesP3MA2017} with memory management via LLAMA) and KOKKOS (\citealt{kokkos9485033} consisting of both core and tool packages). Industry giants are also attempting to tackle the issues of vendor lock-in: for example the Intel oneAPI \citep{costanzo2021early,christgau2020porting} implements specialized APIs as well as the portable SYCL language \citep{alpay2022exploring}. An interesting note is also that HIP which is often viewed as the \mbox{\emph{de facto}} language on AMD hardware, is in fact an abstraction layer which translates commands to the ROCm interface used by AMD hardware, whilst also supporting translation to CUDA on Nvidia hardware. Larger adoptance of HIP remains to be seen, though efforts of supporting HIP on top of oneAPI do exist \citep{HIP-oneapi}.

Whilst utilization of abstraction layers brings with it many benefits, it at the same time detaches the developer from the underlying programming language. Algorithms are restricted to constructions supported by the abstraction layer, which may or may not be useful for the task at hand. Due to the inherent complexity of Vlasiator's memory layout, consisting of a sparse ever-evolving velocity mesh in each spatial cell, and the need to maintain all data in device memory for every solver step, pre-existing abstraction layers were not included in the roadmap of Vlasiator GPU development. After finding directive-based approaches to be too limited, it would have been counterproductive to run into the same issue with portability layers. Abstraction layers are a possible future avenue for fixed-size grids such as the ionospheric and field solver grids, though.

\section{Building semi-Lagrangian Vlasov solvers for CUDA}

With the knowledge of CUDA being a well-established and well-supported GPU programming language with powerful development and profiling tools, and knowing that AMD hardware was to be supported with near-analogous API calls via HIP, Vlasiator was decided to be first ported to CUDA, dubbed \emph{Cudasiator}. Once CUDA development and profiling had sufficiently progressed and HIP support was included, the Cudasiator moniker was dropped in favour of a standard vlasiator\_gpu identification.

\subsection{Unified memory}\label{sec:unifiedmemory}

The Vlasiator simulation code includes several different solvers and routines which act on the modelled particle VDFs. Initialization routines set up the simulation and boundary cell management can re-write cell contents. VDFs are reduced to obtain plasma velocity moments for use by the field solver. Block values are evaluated to find maximum allowed timesteps due to spatial translation. Output routines can process the VDFs to extract partial moments or precipitating energy spectra. Due to all of these accessors, the VDF is stored in unified memory, which can be paged to host or device as necessary. Using unified memory may incur a performance penalty, but reduces the risk of one code section acting on data which might be out-of-date (having been altered by another part of the code) or trying to access inaccessible memory addresses. During early development, small subsections of the solvers were ported one-by-one, allowing verification of correct code behaviour without having to convert all functions to GPU kernels at once. As data movement between the host and the device is much slower than the HBM of GPU hardware, and would cause a significant performance bottleneck, the final state of the Vlasiator code has all VDF-accessing routines run through GPU kernels.

One feature of unified memory on current NVIDIA and AMD architectures is the possibility of page faulting data to the required location. Unified memory is managed as \emph{pages}, large units of data (e.g.\ 4 kilobytes), which reside either in host or device memory. When accessing such managed memory, a lookup to the UM page table is performed, and assuming the data is already in the correct place, it is read from the address indicated by the page. If the memory resides on the host when requested by the device, or vice versa, a \emph{page fault} occurs, and the page lookup table access triggers a memory copy, migrating the page to the requested memory domain. The lookup table is then updated to reflect this new status. Page faults are what allow unified memory to be accessed from both host and device, but repeated page faults cause significant slowdowns as computation must wait for the page transfer to complete. 

Memory can be prefetched by the user if one knows that the data resides in the incorrect location, however, this incurs an overhead of reading the page lookup table and returning only once the potential transfer of each page associated with the buffer in question has been requested. To completely avoid page faults, a synchronization call requesting for this transfer to complete is also required. The combination of these lookup table accesses and GPU API calls will also incur an overhead, so prefetching should only be performed when the software is sure the data resides in the incorrect memory regime. The current GPU version of Vlasiator contains a number of small counters which reside in unified memory, and the page faulting of these back-and-forth is quite detrimental to performance, but manual prefetching is even worse due to the overhead of GPU API calls. If manual prefetching is performed, the GPU driver may be unable to optimize whether a unit of data should be paged back and forth, or if it should simply be copied into a temporary buffer, maintaining the page location where it is.

Of course if a unit of data is not actually required on both host and device, use of unified memory should be avoided. Pure device memory access does not pass through page lookups, whereas host memory usually does. To further improve host performance, in particular when copying data between host and device buffers, the use of \emph{pinned} memory is crucial. Pinned memory is declared outside the paging system, bypassing the need to go via host page lookups when communicating with the GPU device. In Vlasiator, we use pure device buffers and pinned host memory for any single-location temporary buffers which allow it.

\subsection{Re-using device and unified memory buffers}\label{sec:re-using-buffers}

In addition to managing paging of the unified memory subsystem, care must be taken in managing allocations and deallocations. Within the current CUDA and HIP/ROCm ecosystems, managed or unified memory can only be allocated or deallocated when no GPU kernels or memory transfer operations are active. Thus, these are \emph{blocking operations} which can have a significant detrimental effect on performance. Thus, the use of buffers should be designed as to minimise allocations and deallocations, re-using buffers where necessary. This is of particular importance to Vlasiator, where required buffer sizes change constantly as the velocity meshes grow and shrink. Vlasiator thus utilizes a method of maintaining slightly larger buffers than required, re-using them for subsequent kernels, and whenever a re-allocation takes place, making the new allocation larger by a factor of e.g. 30\% in order to postpone the need of future re-allocations. Allocations of pure device memory (or host memory) do not block the whole device, as they can be performed \emph{asynchronously} in one GPU stream, and thus they are not as crucial to avoid.  

\subsection{CPU threads and GPU streams}\label{sec:threads_and_streams}

Vlasiator on CPUs is massively parallel, utilizing MPI for communication between different tasks and supercomputer nodes, OpenMP threading for parallel processing on several CPU cores, and vectorized instructions to process several units of data with a single instruction (SIMD). GPU hardware is also capable of running several operations in parallel, both as GPU warps or wavefronts (SIMT, single instruction multiple threads), but also as several different kernels running in parallel. Vlasiator utilizes this capability in a heterogeneous manner by using several CPU threads to prepare data and queue operations, with the GPU performing the heavy lifting. Each CPU OpenMP thread is given an unique GPU stream to which it can queue operations with each stream acting independent of other streams. Should there be a data dependency between different streams, this can be managed by synchronizing the streams in question, or usually, the whole device, before continuing with enqueueing new GPU kernels or operations.

As each GPU stream will contain e.g.\ the different sub-kernels required by the acceleration Vlasov solver, and any operations queued to that stream will be performed in order, it makes sense to ensure that each stream has its own re-writeable buffers to act on. Thus, on the host side, each CPU thread (controlling a designated GPU stream) maintains its own temporary buffers. These buffers are re-used when that thread (and, by extension, stream) moves on to the next unit of data to be processed, such as the next spatial cell. Due to the use of these per-thread buffers, the translation solver is also launched in several streams, each using the buffer associated with that thread/stream.

%\subsection{Global variables: Velocity mesh parameters, velocity mesh, block container}
%Using \_\_CUDA\_ARCH\_\_ within inlined functions for memory portability.
%Mention that this is not possible with templated classes. Discuss use or relocatable device code during compilation.

\subsection{A Sparse velocity mesh on GPU devices}

In order to be able to access and edit the sparse velocity meshes used in Vlasiator from on-device kernels, a hashmap supporting on-device operations was required. During the porting of Vlasiator to heterogeneous GPU architectures, we performed porting bit-by-bit, requiring thus a hashmap which could be accessed from both host and device, in order to provide correct functionality even before all operations had been completed. For this purpose, we developed \emph{Hashinator}, a portable hashmap library using unified memory and supporting both CUDA and HIP/ROCm. In addition, we made available a portable unified memory vector class called \emph{Splitvector}, on which Hashinator is built. These tools, along with their portability features and excellent performance, are introduced in \cite{Papadakis_2024_Hashinator} and available at \cite{HashinatorZenodo}.

Utilizing the Hashinator and Splitvector libraries, the Vmesh and VBC contents (hashmaps and vectors in dark and pale blue in Figure \ref{fig:Data_constructs}) were converted to unified memory constructs, capable of being accessed from both host and device. Though these intermediate objects themselves could also reside in unified memory, we found that it was preferred to maintain host and device copies of Vmeshes and VBCs, which contain host and device copies of splitvectors and hashmaps, which in turn contain unified memory arrays and counters. This helps to reduce page faulting between host and device, but requires the device copy of the objects to be updated whenever a re-allocation occurs in order to update unified memory pointers. A common occurrence when launching kernels is to evaluate the size of the velocity mesh, which would usually mean a call to the \verb|size()| function of the \verb|globalToLocalMap| hashmap or the \verb|localToGlobalMap| splitvector. We found that carefully caching this value as a host-side integer and updating it whenever the velocity mesh size is adjusted resulted in reduced page faulting.

\subsection{GPU thread parallelism}

As Vlasiator already utilizes memory stored in units of \verb|WID3| floating-point values, acting on 32 (CUDA warp, NVIDIA) or 64 (ROCm wavefront, AMD) elements in one pass might have been relatively straightforward. However, the acceleration solver in particular acts only on one \verb|WID|$\times$\verb|WID| slice at a time, in order to ensure thread-safe storage of phase-space density values. This small slice-by-slice approach is sub-optimal on GPU hardware if the standard value of \verb|WID=4| is retained. A future task will be to refactor the memory access patterns to allow a whole \verb|WID3| block to be processed at once, but for now, a transition to \verb|WID=8| means that each \verb|WID|$\times$\verb|WID| slice has 64 entries and can be efficiently processed on GPU hardware. When a GPU kernel performs operations on a \verb|globalToLocalMap| hashmap, we utilize the warp accessors of Hashinator, doing lookups of a whole warp length at once, further improving thread-parallelism.

\subsection{Porting the Vlasov solvers}

The actual process of porting Vlasiator's Vlasov solvers to GPU kernels is relatively straightforward. Care must be taken to minimize memory transfers between host and device. Re-use of existing buffers is preferred over allocations and de-allocations, which can improve or hinder memory management challenges. 

The acceleration solver has to evaluate the contents of the local velocity mesh and re-assemble it into contiguous memory domains dependant on the Cartesian direction in which the shear in question is taking place. One subsection of this involves converting the velocity mesh GIDs to direction-dependent values $\widetilde{GID}$
(see Section \ref{sec:re-ordering}). The new $\widetilde{GID}$-values are then sorted in order to achieve contiguous input data to the remapping algorithm.
Notably, this is a place where existing sorting libraries such as \Verb|CUB::sort| on CUDA and \Verb|hipCUB::sort| on HIP can be utilized. When evaluating extents of contiguous columns of blocks to be passed to the acceleration solver, thread-parallel GPU search operations provide performance improvements.

The translation solver has a relatively simple flow, allowing merging the preparatory, advection, and storage sections into one large kernel which can be called in parallel utilizing several temporary buffers. The acceleration solver, on the other hand, involves much more transposition and restructuring of data, resulting in a number of small kernels called one after another, which impacts performance by accruing API overhead and sometimes requires passing counter values between the host and the device. After preparatory work, the accelerated cell velocity space must be adjusted to accommodate the new positions in velocity space, which requires calls to the block adjustment algorithms. Furthermore, each spatial cell is accelerated independently of all others, meaning that though they can be processed in parallel, each one requires its own kernel calls in order. Each CPU thread is able to process one cell at a time in its own GPU stream, but extracting acceptable performance from modern GPU hardware may require even further re-factoring of the acceleration solver ordering.

\section{Block adjustment algorithms}

At the heart of Vlasiator's operations, in addition to the SLICE-3D algorithm and the semi-Lagrangian solvers, is the block adjustment algorithm used to maintain the minimum sufficient velocity domain for the problem at hand. A requirement for block adjustment is that in the final state, the velocity mesh should contain the correct GID values, each associated with an unique LID, and the LID-values should be consecutive, i.e. the velocity mesh should be contiguous without any empty spaces left by deleted blocks. In CPU iterations of Vlasiator, the block adjustment algorithm was built using lists of those blocks where content was deemed significant enough, and the use of a \verb|std::unordered_set|, with a serial algorithm to adjust contents. 

\subsection{Serial CPU block adjustment}\label{sec:serial_block}

\begin{figure*}
   \centering
   \includegraphics[width=0.8\textwidth]{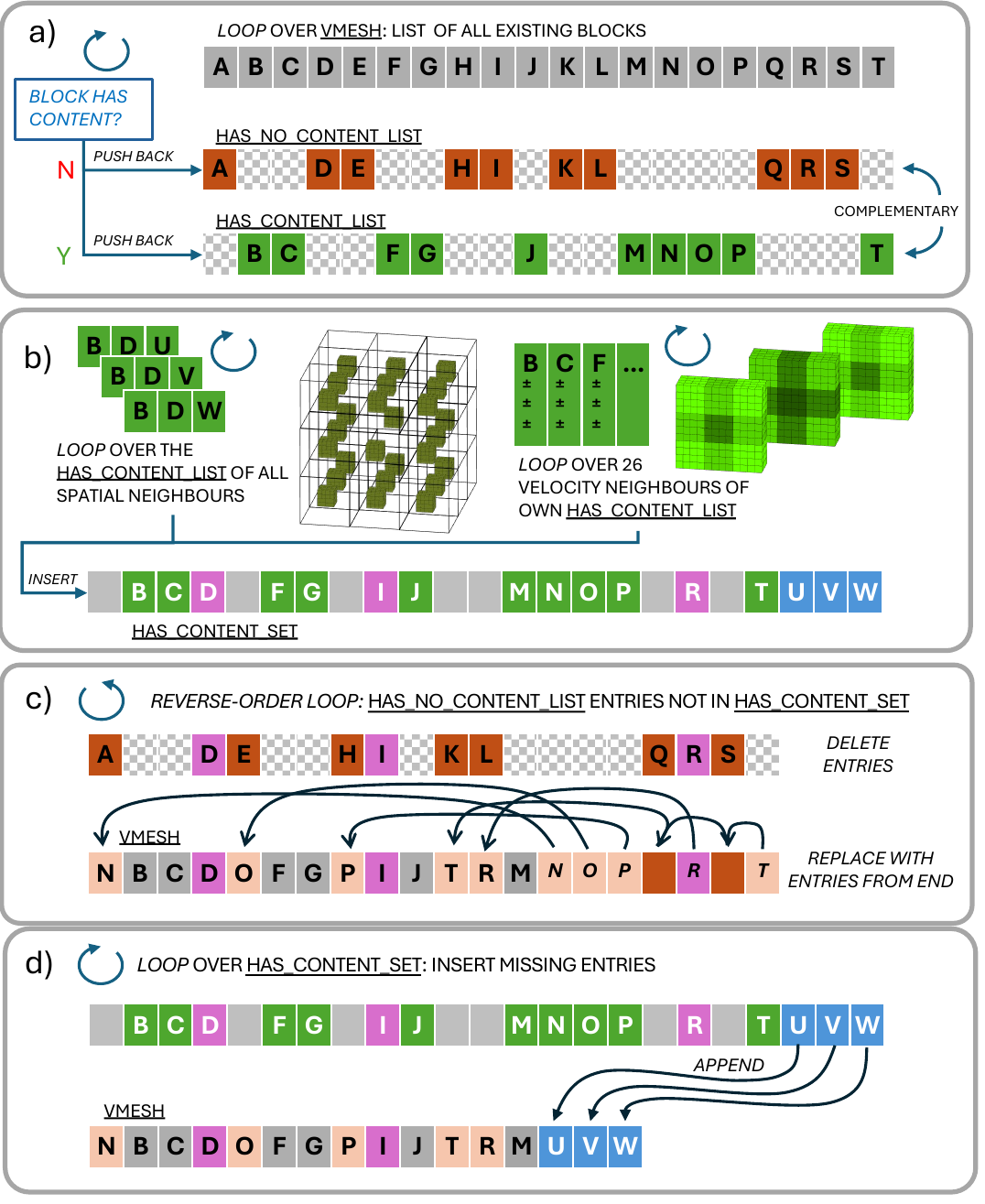}
   \caption{Flowchart of the serial CPU block adjustment algorithm. a) Existing velocity blocks are categorized into two complementary lists (vectors, black lettering, with the checkerboard pattern used for visual spacing of contiguous elements) as either having significant content (green) or not having significant content (red). b) Halo requirements: spatial neighbour cell content list GIDs are added to a \emph{required content} set (white lettering). Local content blocks and their 26 velocity neighbours are also added to this set. Set contents are: blocks with local content (green), blocks with no local content (purple), and blocks not yet existing (blue). c) Local blocks without content are looped over in reverse order. Blocks not found in the requirement set (dark red) are deleted. If the block was not the last entry, it is replaced with the last existing block. d) Loop over the requirement set: appending the velocity mesh with any blocks which do not yet exist (blue).}
   \label{fig:serial_algo}
\end{figure*}

The serial CPU block adjustment algorithm of Vlasiator is described as a flowchart in Figure \ref{fig:serial_algo}. In section a), for each spatial cell, all existing velocity blocks are analyzed. If any of the velocity cells within the block have $f_s >f_\mathrm{th}$, the block is considered a content block. The loop over the block contents can end early once the first value above the threshold is found. Blocks are thus categorized into two complementary lists (vectors, black lettering) as either being classified as content blocks (green) or non-content blocks (red). Section a) must be completed for all spatial cells before moving on. 

Next, in section b), the spatial and velocity halo requirements are fulfilled. For each cell, all its spatial neighbours are identified. For all these spatial neighbour cells, their lists of content cells are looped over, with all contained GIDs added to a \emph{required content} \verb|std::unordered_set| (white lettering). For the local cell in question, any local blocks with content are also looped over, and all these blocks and their 26 velocity neighbours are also added to this set. The set contents are indicated as blocks with content (green), blocks with no content (purple), and non-existing blocks (blue, yet to be added). 

The third phase, section c), involves pruning non-required blocks from the velocity mesh. The list of local blocks without content is looped over in reverse order (that is, starting from the largest GID i.e.\ from towards the end of the allocated buffer). Any blocks which are not found in the required content set (dark red) are deleted from the mesh. If the block was the last entry, the mesh size is simply reduced by one block. However, if the block was not the last entry in the mesh, it is replaced with the contents of the last existing block, and then the mesh is reduced in size by one block. 

In the final phase, section d), a loop is performed over the required content set. Each GID found in the required content set is compared against the current state of the velocity mesh, appending the velocity mesh with any blocks which do not yet exist (blue). This results in the final state, where the velocity mesh contains all required GID values, each associated with an unique LID forming a consecutive list. This serial approach is straightforward, simple, and results in a relatively low amount of data transfer, but is sequential in nature and does not parallelize well on GPU hardware.

\subsection{Parallel GPU block adjustment}\label{sec:parallel_block}

\begin{figure*}
   \centering
   \includegraphics[width=0.7\textwidth]{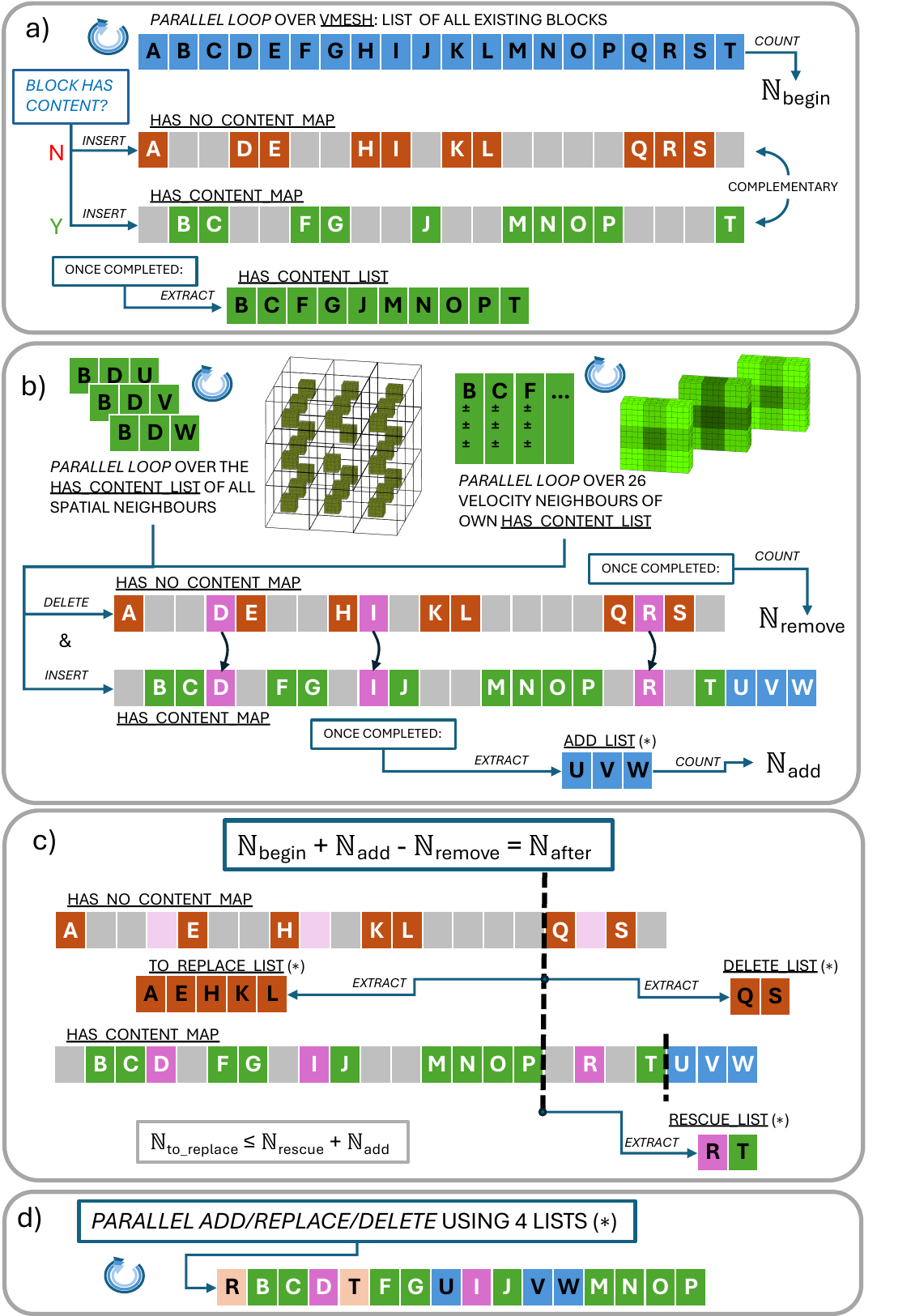}
   \caption{Flowchart of the parallel GPU block adjustment algorithm. a) For each spatial cell, all $\mathbb{N}_\mathrm{begin}$ velocity blocks are divided into two complementary maps (white lettering) as either having significant content (green) or not having significant content (red). The content GIDs are extracted into a list (vector, black lettering). b) Halo requirements: spatial neighbour cell content list GIDs are deleted from the no-content map and added to the content map (purple). Non-existing cells are added to the content map with a identifier tag (blue). Contents are counted, recovering $\mathbb{N}_\mathrm{remove}$ and $\mathbb{N}_\mathrm{add}$ as well as a list of GIDs to add. c) Now with $\mathbb{N}_\mathrm{after}$ known, three lists are parallel-extracted from the maps: replaced, deleted, and rescued blocks. d) First, the mesh is grown, if necessary. Then, blocks are replaced, added, or deleted in parallel, using the four extracted lists. Finally, if necessary, the mesh is shrunk to match $\mathbb{N}_\mathrm{after}$.}
   \label{fig:parallel_algo}
\end{figure*}

The objective of developing the parallel block adjustment algorithm described in this section was to be able to launch GPU kernels which could act in parallel on sections of the same velocity mesh, adding, deleting, adjusting, and moving blocks so that the end result is a valid velocity mesh and so that the kernels do not need to atomically access counters such as taking the next entry from a vector. The algorithm developed for this purpose is displayed in Figure~\ref{fig:parallel_algo}, and it is run for all spatial cells separately.

 In section a), for each spatial cell, all existing velocity blocks (within the \verb|localToGlobalMap| list or vector, black lettering) are analyzed. The original count $\mathbb{N}_\mathrm{begin}$ is stored. A kernel examines all blocks in parallel, and for each block an in-kernel thread-parallel reduction analyzes all contained phase-space cells at once. If any phase-space cell fulfills $f_s > f_\mathrm{th}$, the block is considered a content block. Blocks are thus categorized into two complementary maps (white lettering) as either content blocks (green) or non-content blocks (red). The maps store the block GID as the key and the LID as the value. After the parallel insertions, the entries from the content map are extracted into a contiguous \emph{content list} (vector, black lettering). Section a) must be completed for all spatial cells before moving on. 

Following the gathering of content lists, section b) ensures all halo requirements are fulfilled. For each cell, all its spatial neighbours are identified. For all these spatial neighbour cells (managed in parallel), their content lists are looped over in parallel, and the GIDs are evaluated against the local content maps. Any neighbour content GIDs found in the no-content map are deleted from there, and instead placed in the content map (purple). If the neighbour GID does not yet exist in either map, it is added to the content map with an invalid LID identifier flagging it as a cell yet to be added (blue). For the local cell, all content blocks and their 26 velocity neighbours are also looped over in parallel, and values are added to the content map and deleted from the no-content map as before. As listed, all of these operations can be performed in parallel. The content map elements are indicated in the figure as blocks with local content (green), blocks with no local content (purple), and non-existing blocks (blue, yet to be added). After all halo requirements have been evaluated, the two resultant maps are evaluated in order to calculate the size of the velocity mesh after all required operations will be complete. The size of the no-content map provides the number of blocks to remove $\mathbb{N}_\mathrm{remove}$. The content map is passed to a parallel stream compaction routine, extracting all entries flagged as yet-to-be-added GIDs into a list of blocks to add, and the size of this list provides $\mathbb{N}_\mathrm{add}$.

Next, knowing that \mbox{$\mathbb{N}_\mathrm{begin} + \mathbb{N}_\mathrm{add} - \mathbb{N}_\mathrm{remove} = \mathbb{N}_\mathrm{after}$}, the generated content and no-content maps can be evaluated in order to prepare lists of GIDs for parallel processing. Using $\mathbb{N}_\mathrm{after}$ as an extraction criterion, checking if map entries have a valid value (LID) less than it, we use parallel \emph{stream compaction} to extract three lists of blocks. Any blocks with content residing beyond the final size of the mesh are designated blocks to rescue. Blocks without content residing beyond the final size of the mesh are to be simply deleted. Blocks without content within the resultant domain of the mesh are to be removed and replaced with either added or rescued blocks. 

In the final step, shown in section d) the four constructed lists (blocks to be added, blocks to be deleted, blocks to be replaced, and blocks to be rescued) are used in a parallel operation altering the actual velocity mesh associated with the spatial cell in question. If \mbox{$\mathbb{N}_\mathrm{begin} < \mathbb{N}_\mathrm{after}$}, the size of the mesh is increased, in order to have sufficient space for all parallel operations. This size adjustment is done on-device if possible in order to avoid page faulting of counters. If the capacity of the map is insufficient, it is performed on host instead. Then, a GPU kernel is launched, where each parallel instance operates on one value or pair of values from the generated lists. Depending on the relative sizes of the four lists, the operation can involve outright deletion of a block, addition and emptying of a new block, replacing an existing block with a new empty block, or replacing an existing block with a block to be rescued. After all these parallel operations have been completed, the contents of the velocity mesh are in their final state. Then, if \mbox{$\mathbb{N}_\mathrm{begin} > \mathbb{N}_\mathrm{after}$}, the size of the mesh is decreased to match the contents.

\section{Data reduction methods}

The Vlasov solvers of Vlasiator act as a 6-dimensional advection equation, but the Maxwell-Vlasov system also requires input from electromagnetic fields which are propagated by the field solvers. Those field solvers take plasma velocity moments as input, necessitating an efficient way to reduce VDF contents.

\subsection{Time step constraints}

Although Vlasiator's semi-Lagrangian solvers do not place a CFL condition as such, a similar limitation is in place due to a limited thickness of the spatial ghost cell layer. Thus, the dynamic simulation time step may be constrained by the largest advection velocity found in the simulation. To this end, a timestep limit reduction kernel is required, which finds the smallest value of $\Delta x / v_x$, $\Delta y/v_y$, or $\Delta z/v_z$ in the simulation, which, with a safety margin, gives the largest allowed spatial advection time step.

\subsection{Plasma moments}\label{sec:moments}

Reduction of plasma velocity moments \citep[see, e.g.,][]{Palmroth_2018_LRCA} for the simulation domain is performed by gathering device-side pointers to the velocity mesh and block container structures for all local spatial cells, preparing output buffers, and then calling a massively parallel reduction kernel over all local spatial cells. Earlier versions of Vlasiator called a reduction kernel on one cell at a time, but the overhead from repeated kernel calls and back-and-forth transfer of minuscule units of data was prohibitively slow. Kernel merging with no other changes was found to provide a speed improvement of order $10\times$.

\subsection{Output data reducers}

A single full state of the Vlasiator simulation, as seen in a simulation checkpoint file, can be several terabytes. In order to facilitate data analysis with more reasonable file sizes, output files with reduced data are generated at regular intervals. The most commonly used values stored in these files are values such as electric and magnetic field vectors and the aforementioned plasma moments. However, more detailed data reducers operating over the velocity space of each spatial cell, such as measures of non-Maxwellianity or sub-region analysis (particle precipitation, moments of supra-thermal sub-populations) are also available. Vlasiator implements a flexible ARCH-lambda approach for data reduction, where the calculations to be performed in order to extract the necessary information are prepared as a C++ lambda function, which is passed over to an interface which then either passes it to CPU-based loops, or if GPU hardware is being used, constructs a kernel to which it is passed. The lambda approach as such does not accrue noteworthy overhead, but if the reduction kernel is called for each spatial cell in turn, API overhead as described in Section \ref{sec:moments} can become prohibitive.

\section{Unified cross-platform codebase}

A common issue with large software packages which are developed by several people in parallel for several different platforms is that of code stagnation and cross-compilation issues. When developing Vlasiator support for GPU architectures, a certain amount of code duplication was unavoidable. However, a strong object-oriented approach allows selective compilation of only those source code files which contain objects and routines to be used by the target architecture. The Vlasiator GPU version is thus developed as one unified codebase, with the compilation target (CPU, CUDA, or HIP/ROCm) chosen by environment variables at compile time.

%\subsection{CUDA and HIP/ROCm with macros}

Early during development planning, a joint CUDA/HIP development approach was chosen. Initially, Vlasiator was developed as a version called Cudasiator, with direct CUDA instructions, libraries and classes. The initial plan was to use automatic conversion scripts (such as \verb|hipify-perl.sh|, included as part of HIP/ROCm) during compile time to convert the source code to a suitable form. This approach has a significant limitation, in that any code development performed on the HIP/ROCm platform would then need to be reverse-engineered back into CUDA code before being committed to the code repository.

Joint support for AMD hardware with the HIP language was subsequently achieved by converting all Vlasiator GPU API calls to a dedicated macro naming scheme. Depending on compile-time instructions, these macros are then converted by the compiler preprocessor into either CUDA or HIP calls. For example, the Vlasiator codebase will contain a call to \verb|gpuStreamSynchronize()|, which will be converted to either \verb|cudaStreamSynchronize()| or \verb|hipStreamSynchronize()|, as required by the architecture. All these macro definitions are collected into two header files, and the one to be included in compilation is defined via compile-time environment selection. As CUDA and HIP/ROCm are mostly interoperable, this process is relatively straightforward. Special cases include specialized kernels which must know the size of a warp or wavefront, which is also declared as a macro \verb|GPUTHREADS| which resolves to either 32 or 64. 

We note that this approach is suitable also for situations where the CUDA and HIP interfaces differ, such as in-kernel ballot operations. A call to \verb|gpuKernelBallot(mask, input)| will be pre-processed into either \verb|__ballot_sync(mask, input)| (CUDA) or \verb|__ballot(input)| (HIP), ensuring correct compilation and operation. However, this has the limitation that the user code must always use a full mask, somewhat limiting the capabilities of CUDA operation. At the same time, advanced CUDA features such as cooperative groups are also not supported as they have no direct HIP counterpart \citep{tsai2021}.

\section{Performance results}\label{sec:performance}

Efficient use of GPU hardware requires often considerable refactoring of code and development of new algorithms. Expecting a freshly ported code to match or exceed the performance of a CPU code which has been developed and optimized over several years or even a decade is not necessarily realistic. Still, GPU hardware provides access to high-bandwidth memory and tens of thousands of compute cores per device, so some level of performance improvement might be expected.

To showcase the capabilities of Vlasiator, we present here simulations performed on the Mahti supercomputer at CSC in Kajaani, Finland. A single Mahti node contains two AMD Rome 7H12 CPUs with 64 cores each. The CPUs are based on AMD Zen 2 architecture, supporting the AVX2 vector instruction set, and run at 2.6 GHz base frequency (max boost up to 3.3 GHz). GPU nodes are equipped with four Nvidia Ampere A100 GPUs with 40 GiB of memory each. For the purposes of this comparison, we thus select 32 cores from one CPU, with no multithreading active, which matches the capabilities assigned to each A100 GPU. The CPU tests utilize CPU instructions only, with AVX2 vectorization, whereas the GPU tests operate on one A100 device from 32 concurrent CPU threads and 32 CUDA streams. The GPU versions were compiled with NVCC in an environment with gcc version 10.4.0 and CUDA version 12.1.1. Compilation of HIP code for CUDA \citep[as shown in, e.g.,][]{tsai2021} was excluded from this test. The CPU versions were compiled with gcc version 11.2.0 and utilize the jemalloc version 5.2.1 memory manager \citep{jemalloc}.

Furthermore, we include some results from simulations performed on the LUMI-G supercomputer at CSC in Kajaani, Finland. A single LUMI-G node contains one AMD Trento 7A53 64-Core Processor and four AMD Instinct MI250x GPUs each with 128 GiB of memory. Each MI250x GPU consists of two independent Graphics Compute Dies (GCDs), visible to the user as separate GPUs despite residing on the same hardware card. The recommended configuration is to use 6 CPU threads per GCD, leaving two threads per GCD for system operations. The HIP/ROCm version was compiled with the \verb|hipcc| compiler wrapper, HIP version 5.6.3, AMD clang version 16.0.0, and ROCm version 5.6.1.

\subsection{Simulation setup}\label{sec:khi}

\begin{figure}[htp]
    \includegraphics[width=\textwidth]{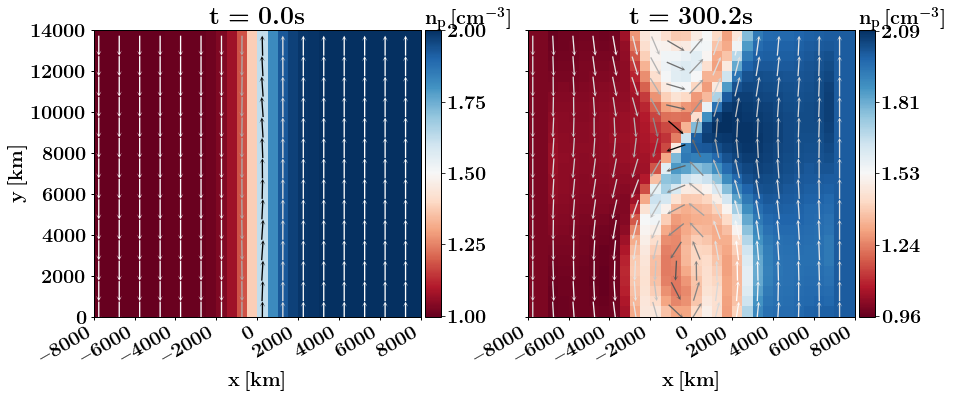}
    \caption{Initial (left) and intermediate (right) states of a sample simulation of magnetized plasma at a shear layer where a Kelvin-Helmholtz instability develops. The presented small simulation was run for a total of 700\,s in order to evaluate performance metrics from the different code versions. The colour map indicates plasma density with the two sides initially in pressure balance, with the overlaid vectors indicating flow direction. The vector colours indicate flow magnitude with white vectors indicating high speeds and dark colors corresponding with low speeds. A small initial velocity perturbation can be seen in the vectors of panel a.}
    \label{fig:KHI}
\end{figure}

The simulation test case consists of a plasma shear flow interface, with two counter-propagating domains of magnetized plasma. The simulation domain is compact and 2-dimensional, spanning 32 cells in the $X$-direction and 28 cells in the $Y$-direction with cell edges of extent 500 km. In the left ($-X$) half of the simulation domain, plasma has a density of $1\,\mathrm{cm}^{-3}$ and a bulk flow velocity of $V=(0,-187.5,0)\,\mathrm{km\,s}^{-1}$. In the right ($+X$) half of the simulation domain, plasma has a density of $2\,\mathrm{cm}^{-3}$ and a bulk flow velocity of $V=(0,+187.5,0)\,\mathrm{km\,s}^{-1}$. The magnetic field is uniform with a value of $B_z=12\,\mathrm{nT}$, resulting in an ion gyroperiod of $5.4268\,\mathrm{s}$. The domains are set to have a pressure balance, with a uniform pressure of $22.3\,\mathrm{nPa}$, resulting in temperatures of $384\,\mathrm{MK}$ in the left half and $192\,\mathrm{MK}$ in the right half of the simulation domain. The central transition follows a \verb|tanh| profile with a transition width of $1000\,\mathrm{km}$. In addition, a small velocity perturbation of amplitude $3.5\,\mathrm{km\,s}^{-1}$ is superposed, with a wavelength spanning the simulation box in the $Y$-direction. In each spatial cell, the velocity space is constructed with maximal extents of $\pm 3000\,\mathrm{km\,s}^{-1}$ in each direction and a velocity resolution of $12.5\,\mathrm{km\,s}^{-1}$ in all of $\Delta v_x$, $\Delta v_y$, and  $\Delta v_z$. Figure \ref{fig:KHI} shows the simulation states at initialization and after $300.2$\,s of simulation, showing how a Kelvin-Helmholtz instability \citep{Tarvus2024} develops causing uneven plasma mixing at the interface. The simulations are run for a total of $700$\,s which corresponds to almost 129 gyroperiods, with the KH instability growth saturating after about 50 gyroperiods \citep[as shown in Figure 4 of][]{Tarvus2024}.  This simulation time results in 2337 timesteps when using a value of \verb|WID=4|, and 2494 timesteps when using a value of \verb|WID=8|. This difference stems from \verb|WID=8| resulting in a stricter CFL translation constraint due to larger velocity space extents, which in turn arise from the large ghost halo domains, as exemplified in Figure \ref{fig:wid4to8}. We note that this simulation usefully has a stable computational cost (average block counts remain roughly constant) throughout whilst still involving two different plasma domains and showcasing the relevant dynamics.

\subsection{Performance comparison on NVIDIA hardware}

Both CPU and CUDA versions of Vlasiator were used to run the aforementioned test simulation. For both architectures, variants with both \verb|WID=4| and \verb|WID=8| were compiled and simulated. The \verb|WID=4| versions resulted in approximately 52M blocks in the simulation in total, so roughly 80K blocks per spatial cell. This is greater than usually seen in Vlasiator simulations, but a realistic count in e.g.\ hot magnetosheath portions of 2D-3V magnetospheric simulations. By comparison, the \verb|WID=8| versions resulted in approximately 8.8M blocks in the simulation in total, so roughly 12.7K blocks per spatial cell. The ratio is not quite $8\times$ due to the larger ghost halo domains, as Figure \ref{fig:wid4to8} would suggest.

\begin{figure}[htp]
    \includegraphics[width=0.8\textwidth]{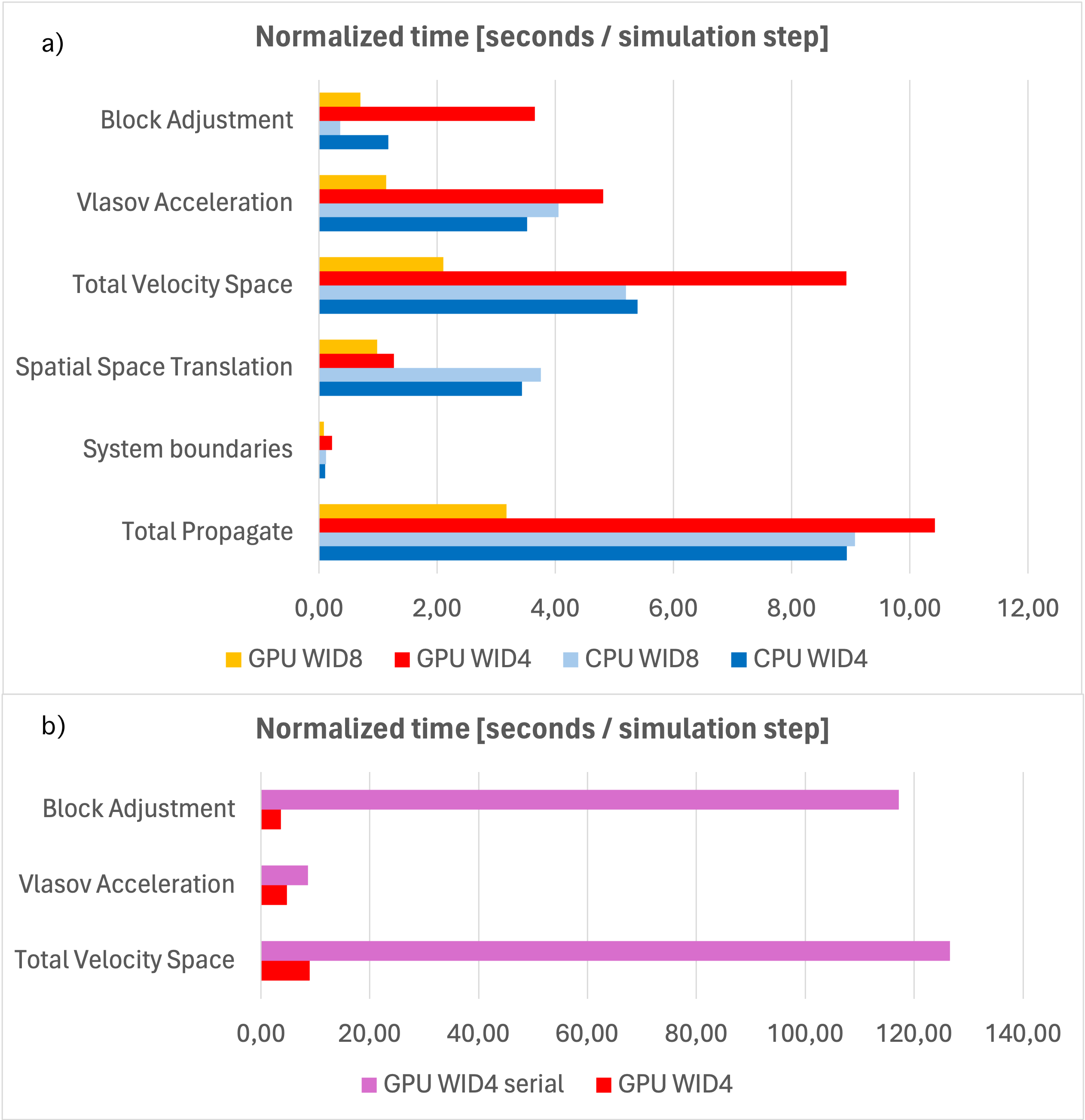}
    \caption{Normalized performance in seconds per simulation step for different compilation variants of Vlasiator, for the Kelvin-Helmholtz simulation case presented in Section \ref{sec:khi} and Figure \ref{fig:KHI}. Simulations are run on 32 CPU cores and where indicated, one NVIDIA A100 card. Smaller values are better. Panel a): Performance of solver portions for Block adjustment and Vlasov Acceleration, the two main components of Total Velocity Space, as well as Spatial Space Translation and System Boundaries updates, which together amount to most of the time spent in Total Propagation. Red and orange bars indicate GPU performance and light and dark blue bars indicate CPU performance. Light colours correspond to values of \protect\UseVerb{WID=4}, dark colours to \protect\UseVerb{WID=8}. Panel b): For comparison, performance of the GPU version utilizing a direct port of the serial CPU block adjustment algorithm is shown for the velocity space propagation, for \protect\UseVerb{WID=4} only.}
    \label{fig:KHI_barchart}
\end{figure}

The bar charts in panel a) of Figure \ref{fig:KHI_barchart} show that for CPU variants, most solver portions perform roughly at similar efficiency for both \verb|WID=4| and \verb|WID=8|. A notable exception is block adjustment, where \verb|WID=8| takes the lead due to much smaller block lists to manage. The GPU version with \verb|WID=4| performs overall somewhat weaker than the CPU versions, due to slightly less efficient block adjustment performance. The total time spent in velocity space updates is larger than CPU variants by almost 80\%, which is partly due to the included block adjustment routines, and partly due to insufficient optimization of the small auxiliary kernels of acceleration. With \verb|WID=8|, however, the GPU version takes a strong lead by utilizing GPU thread parallelism to its fullest. Total propagation time is improved over CPU variants by a factor of three, and both GPU variants show a factor of $>3$ improvement also in spatial translation.

As an additional point of comparison, the simulation was attempted with the GPU version which had the CPU-based serial block adjustment (see Section \ref{sec:serial_block}) implemented in place of the new parallel GPU block adjustment algorithm (see Section \ref{sec:parallel_block}). This version did however perform the computations on-device in order to reduce page faulting. Nevertheless, this serial approach was much slower than any other version, as shown in panel b) of Figure \ref{fig:KHI_barchart}.

These results showcase that with careful design and algorithmic improvements, the capabilities of modern GPU hardware can be harnessed to enhance performance of even complicated simulation simulation codes such as Vlasiator with ever-changing meshes. However, the test case presented here involves complex velocity meshes with large block counts and a relatively small 2-D spatial grid. A similar performance improvement should not be expected in all cases. Preliminary testing with several MPI tasks (not shown) did not yet result in acceptable performance, which we attribute to a lack of GPUdirect RDMA implementation, forcing all MPI communication to go via the host. This hurdle will be tackled in future updates, as efficient multi-node GPU implementations have been shown to be possible \citep{PEKKILA2022102904}.

\subsubsection{Roofline analysis on NVIDIA hardware}

\begin{figure}[htp]
    \includegraphics[width=0.85\textwidth]{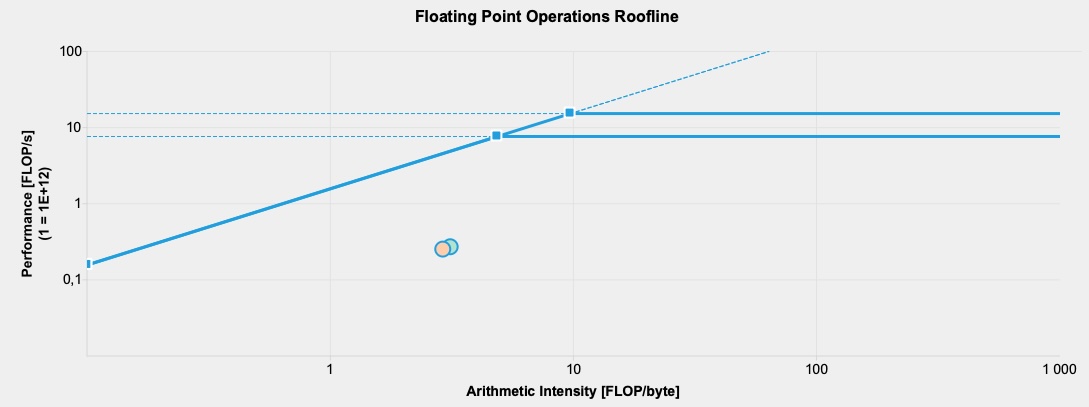}
    \caption{Roofline analysis of the Vlasiator spatial translation kernel on an NVIDIA A100 GPU card. The pale orange dot indicates achieved double precision floating point performance (246 GFLOP/s, arithmetic intensity 2.93), and the pale green dot indicates achieved single precision floating point performance (264 GFLOP/s, arithmetic intensity 3.14). Future improvements such as instruction merging and optimizing memory access patterns are expected to improve performance by up to an order of magnitude.}
    \label{fig:trans_roofline}
\end{figure}

In order to gauge the expected future performance improvements which can be coaxed out of Vlasiator's GPU version, we used NVIDIA's performance analysis tools to, amongst other things, complete \emph{roofline analysis} of the Vlasiator spatial translation kernel, as shown in Figure \ref{fig:trans_roofline}. Roofline analysis evaluates the achieved memory transfer rates and the number of arithmetic operations performed by the kernel in relation to the maximum available memory bandwidth and compute capability. The $X$-axis of a roofline plot is the arithmetic intensity, or ratio of arithmetic operations to bytes accessed. The left-hand side of the plot shows a slope, indicating maximum possible performance as limited by available memory bandwidth. On the right-hand side of a roofline plot operations are limited by arithmetic compute capacity. The Vlasiator translation kernel contains both double and single precision floating point operations and has an arithmetic intensity of roughly 3, placing it in the memory-bound regime. The $Y$-axis indicates actual performance in floating point operations per second (FLOP/s) with the Vlasiator translation kernel achieving on the order of 250 GFLOP/s. 

We note that this roofline analysis was somewhat artificial due to constraints placed by the NVIDIA CUDA toolchain. Instead of using the usual per-thread per-stream temporary buffers as mentioned in Section \ref{sec:threads_and_streams}, this test was performed with only a single thread and stream. The temporary buffer of this thread/stream was ensured to be sufficiently large to allow the whole translation task to fit in memory at once. Even so, analysis of this profiling indicated that the kernel experienced stalling during memory reads, explaining why the achieved performance is roughly an order of magnitude below the maximum achievable value. A significant portion of these stalls were \emph{long scoreboard stalls} when loading VDFs from global accelerator memory. Future optimizations to memory access patterns and pre-caching may improve the performance of this section.

\subsubsection{Overall performance on CUDA and NVIDIA hardware}

The performance of the acceleration portion of Vlasiator's solvers on GPU hardware is less straightforward to evaluate, as it consists of several small kernels to be operated in succession. Additionally, the acceleration process updates spatial space for one cell at a time (per thread and stream), so filling the whole device with one kernel invocation is not possible, barring a comprehensive re-write of the whole acceleration framework. This may be a task for a future code update.

\begin{figure}[htp]
    \includegraphics[width=\textwidth]{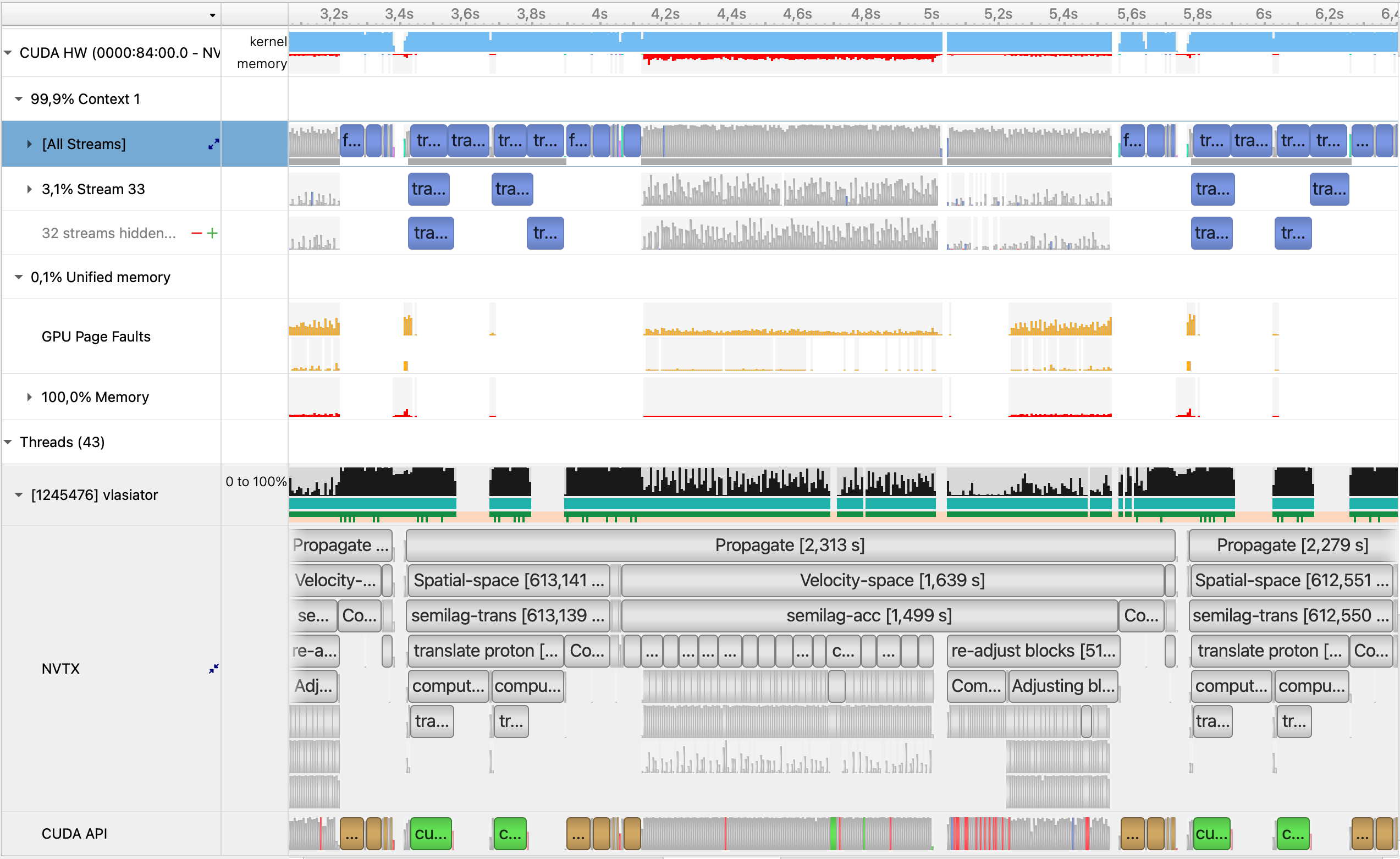}
    \caption{NVIDIA Nsight Systems timeline view of a Vlasiator GPU simulation over one timestep, compiled with \UseVerb{WID=8}. Nested NVTX ranges are triggered by user-provided calls in the source code. The spatial space (translation) solver is able to get away with relatively few kernel invocations with high occupation and only very few page faults occurring between the kernels. The velocity space propagation (acceleration), conversely, consists of a multitude of small kernel invocations with frequent page faults of the unified memory subsystem causing stalling. Notably, the first section of block adjustment (``Com..." within the ``re-adjust blocks" timer range) consists of computing content lists, and has already been optimized to minimize page faults. Nevertheless, the large amount of small kernels still ends up being relatively inefficient. Performance is also hindered whenever GPU page faulting (orange indicators) occurs.}
    \label{fig:CUDAprofile}
\end{figure}

In addition to per-kernel performance, total Vlasiator GPU performance was found to be impacted considerably by page faulting and GPU API calls (see Section \ref{sec:memory_api}) due to several small kernel invocations. This is exemplified in Figure \ref{fig:CUDAprofile}, which shows a timeline over one simulation timestep for the GPU \verb|WID=8| run as visualized via NVIDIA Nsight Systems, for the previously presented KHI simulation. We utilize the phiprof \citep{phiprof} profiling library, which involves user-defined nested timer regions building up to a profile tree. When compiled on a GPU-compatible platform, phiprof can leverage NVTX (NVIDIA) or ROC-TX (AMD) tracing hooks, allowing the same nested timer ranges to show up in GPU-specific profiling tools.

As an example, the first section of block adjustment (``Com..." within the ``re-adjust blocks" timer range) in Figure \ref{fig:CUDAprofile} consists of computing content lists for all spatial cells, a task which has already been optimized to minimize page faults. Before this optimization, page faulting of small counters resulted in spending roughly $3\times$ the amount of time in this range. A future optimization task is to consolidate content list updates into a few large kernel launches which is expected to provide an up to $10\times$ performance improvement, based on the results of Section \ref{sec:moments}.

\subsection{Performance on HIP/ROCm and AMD hardware}

\begin{figure}[htp]
    \includegraphics[width=0.85\textwidth]{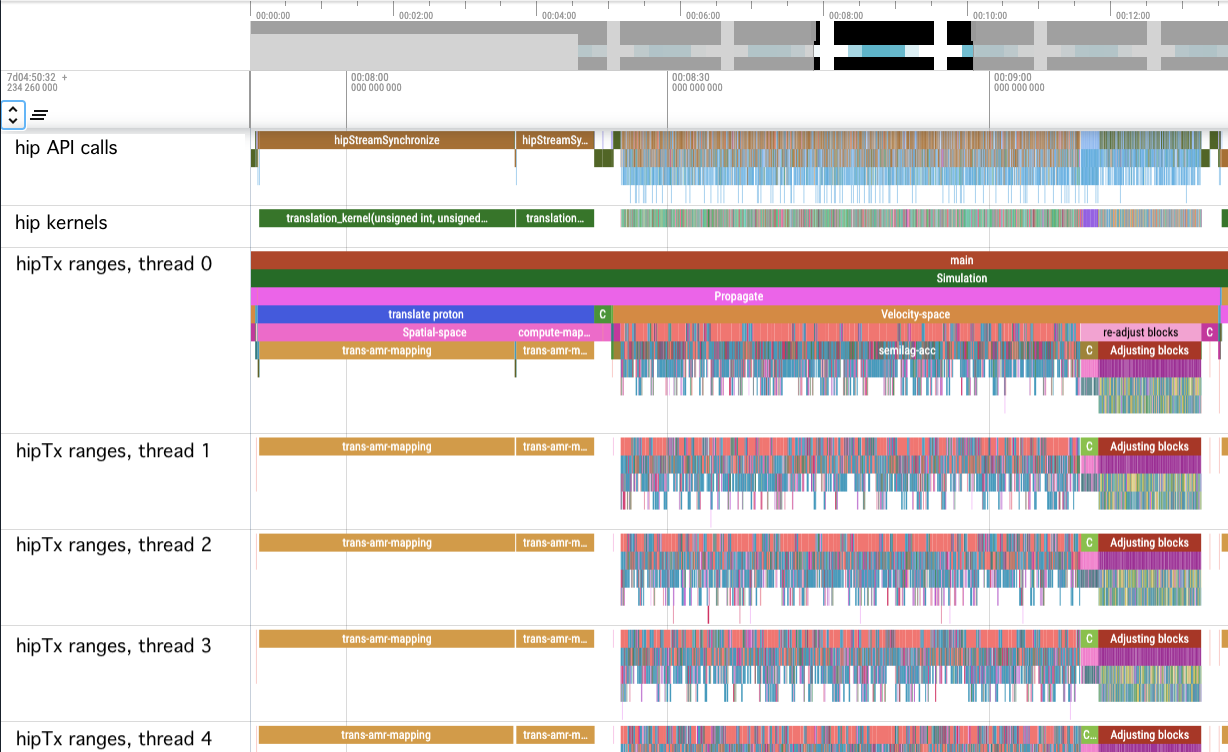}
    \caption{ROCm \protect\UseVerb{rocprof} Trace of Vlasiator GPU solvers on LUMI-G (AMD MI250X accelerator). The relative time spent in various solver tasks differs only slightly from the CUDA version shown in Figure \ref{fig:CUDAprofile}. Visible at the top is the timeline, followed by HIP API calls and kernel launches (overlapping from several GPU streams). Below, we show nested roctx ranges which are triggered by user-provided calls in the source code, showcasing time spent in different parts of the solvers. As shown here, translation (blue range on the left) and moment reduction (letter C, green background next to translation) ranges involve very few wide kernels, whereas velocity space updates involve excessive simultaneous API calls and kernel launches.}
    \label{fig:hip_trace}
\end{figure}

%And discussion of differences in atomic operations, coarse vs fine-grained memory

To evaluate the performance of Vlasiator on AMD hardware, and to validate the usefulness of the unified codebase approach of Vlasiator, we also ran the same Kelvin-Helmholtz simulation on LUMI-G utilizing HIP/ROCm and AMD hardware. To speed up access to unified memory allocations on LUMI-G, we pre-loaded a custom memory manager (\url{https://github.com/sfantao/vlasiator-mempool}). Similar to Figure \ref{fig:CUDAprofile}, Figure \ref{fig:hip_trace} shows a Vlasiator timeline over one simulation time step, showing the different solver sections at work. Comparing the two timelines confirms that overall, the two timelines appear similar, with semi-Lagrangian acceleration taking the longest, and block adjustment and translation taking a quarter to a third of the overall time. On LUMI-G, the translation kernel appears somewhat slower relative to acceleration. This was initially attributed to the low amount of hardware threads resulting in less temporary buffers available for the translation kernel, constraining parallel launch parameters, but increasing the buffer sizes did not appear to alleviate the issue.

\begin{figure}[htp]
    \includegraphics[width=0.8\textwidth]{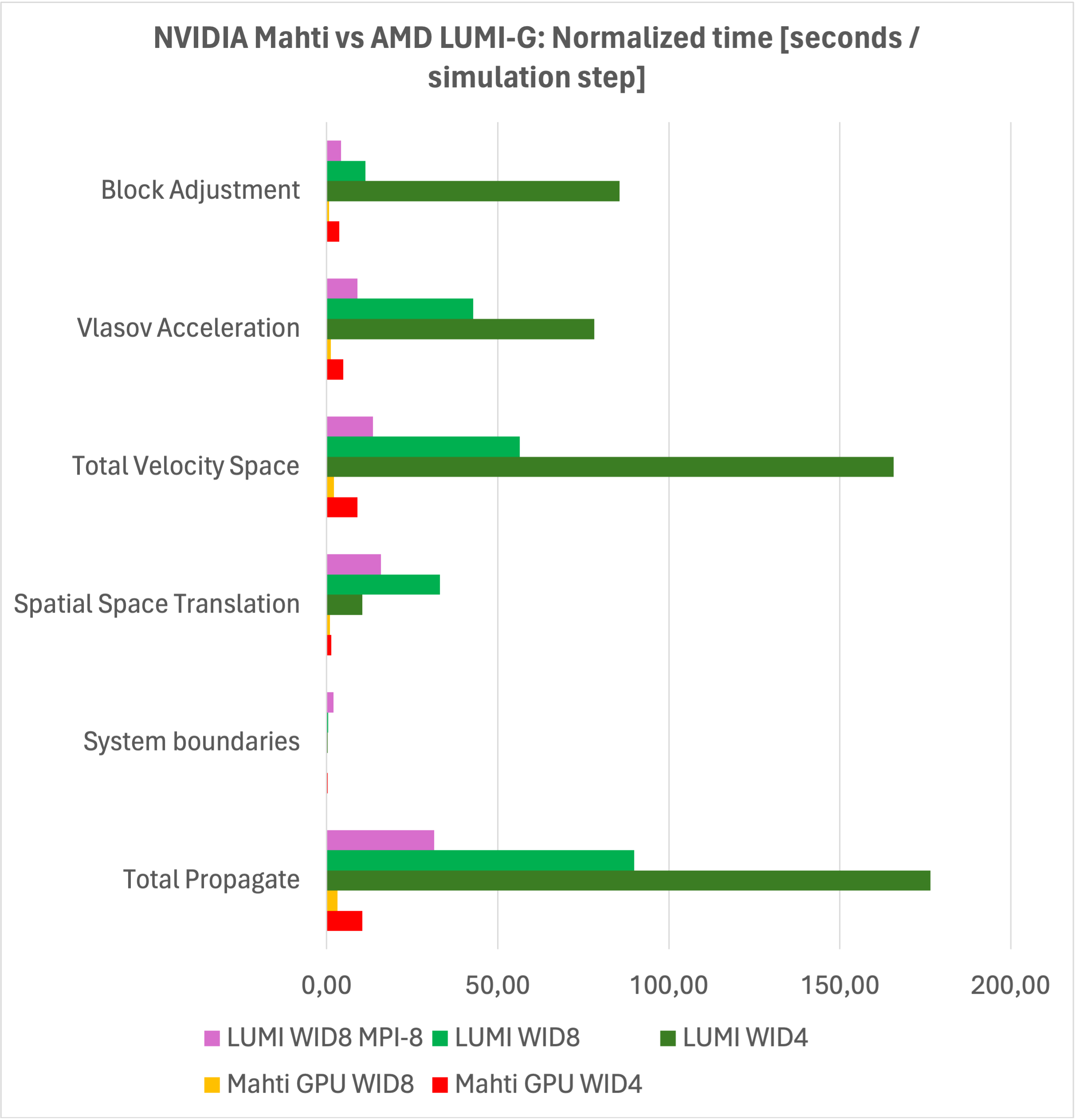}
    \caption{Normalized performance in seconds per simulation step for different compilation variants of Vlasiator, for the Kelvin-Helmholtz simulation case presented in Section \ref{sec:khi} and Figure \ref{fig:KHI}. Smaller values are better. This graph shows performance on LUMI-G using AMD MI250x hardware, with reference values using Mahti hardware equivalent to those in Figure \ref{fig:KHI_barchart}. The LUMI-G runs are run with 6 CPU threads and one GCD from a MI250x GPU, except for the run indicated MPI-8 where 8 MPI tasks were launched, each one controlling one GCD with 6 CPU threads. As can be seen, AMD performance is not yet at an acceptable level, but support for GPU-aware MPI allows for several MPI tasks to be utilized without undue performance penalty.}
    \label{fig:KHI_barchart_AMD}
\end{figure}

This slight relative change in translation versus acceleration is less surprising than the overall poor performance on the tested hardware, as shown in Figure \ref{fig:KHI_barchart_AMD}. Despite driving a powerful accelerator with several CPU cores and hardware threads, LUMI-G performance was roughly 3.7\% of NVIDIA performance with \verb|WID=8| and 5.9\% with \verb|WID=4|. A single AMD MI250x GCD is not expected to match the performance of a single NVIDIA A100 card, but they should be able to achieve the same magnitude of performance. Even the optimized straightforward reduction kernels used for velocity moments of the VDFs (see Section \ref{sec:moments}) showed a $50\times$ penalty in performance. This may be partially due to device drivers in need of updating (which is the responsibility of system administrators, not end users) or, for example, slower response time in the unified memory paging subsystem (see Section \ref{sec:unifiedmemory}). A positive note is that running several MPI tasks was able to benefit from the increased throughput, with the LUMI-G system supporting direct network access from GPU memory (GPUDirect RDMA). The apparent strong scaling, going from 1 MPI task to 8 tasks, is not representative of overall scaling as a single MPI task will be able to skip all network communication entirely. Additionally, the presented simulation case was chosen to have large block counts, which results in heavier-than-usual MPI communication patterns. Multi-node GPU performance analysis will be the topic of a subsequent study.

\section{Conclusions and outlook}

In this paper, we have presented evidence of how porting a complicated modern simulation software to support heterogeneous CPU-GPU architectures whilst maintaining portability and updating a sparse velocity mesh can be challenging. Through algorithmic re-design and memory management optimization, a significant portion of the potential of GPU hardware can be harnessed, as is shown by the results in Figure \ref{fig:KHI_barchart}. The use of unified memory (see Section \ref{sec:unifiedmemory}) has been very helpful during the porting process, but places certain constraints on performance due to the overhead of page faulting. A future task will be to use a managed memory pool library \citep[such as UMPIRE][]{umpire} for mitigating blocking operations such as (de)allocations (see Section \ref{sec:re-using-buffers}).

The impressive performance achieved in e.g.\ moments reductions when using a single massively parallel kernel showcase that transitioning from several concurrent or consecutive small kernel launches to few large launches is crucial to achieving optimal performance on systems with NVIDIA hardware. This task has been mostly completed for the Vlasiator spatial translation solver, but applying the same formalism to the acceleration solver will be a major undertaking.

Our results show that CUDA and HIP/ROCm are indeed sufficiently interchangeable. Building a unified codebase which is translated to the language in question using preprocessor macros is quite feasible and we would recommended such an approach. The underwhelming performance achieved on LUMI-G and AMD hardware is yet to be explained, though extensive use of unified memory is one potential culprit. This investigation is thus ongoing.

Porting modern HPC simulation softwares to support GPU hardware is not a question of \emph{IF}, but rather a question of \emph{WHEN} and \emph{HOW}. We conclude that due to the extent of required algorithmic changes, the task should be initiated as early as possible, and we advise use of industry-standard methods with good compiler and profiling support. A report on further GPU performance improvements of Vlasiator, and in particular, use of Vlasiator in multi-GPU multi-node environtments, is left as topic of a subsequent study.

\section*{Data Availability Statement}
% Code availability on GitHub
% Data availability
The Vlasiator simulation code is distributed under the GPL-2 open source license at \url{https://github.com/fmihpc/vlasiator} \citep{vlasiatorcode} with the GPU branch pre-release available at \cite{vlasiatorgpucode}.
%Full simulation data for the presented analysis is stored in the University of Helsinki Datacloud. Data presented in this paper can be accessed by following the data policy on the Vlasiator website \url{https://helsinki.fi/vlasiator}. 
Data handling and visualisation is supported through the Analysator python package \url{https://github.com/fmihpc/analysator} as well as a plugin for the VisIt visualisation system \citep{HPV:VisIt}.
Configuration files defining simulation parameters for runs presented in this paper are available from the authors upon request. %Simulation config files and raw performance data are available for download from reference \onlinecite{idaGPUdata}.

\section*{Acknowledgements}
This work builds upon the results achieved through the European Research Council Consolidator Grant 682068-PRESTISSIMO. The majority of the results presented in this paper were achieved with funding provided by the Research Council of Finland project ICT\_SUNVAC, grant number 335554. Further support was granted by the EuroHPC ``Plasma-PEPSC'' Centre of Excellence (Grant number 4100455) and the Research Council of Finland matching funding (grant number 359806). 
% Give acknowledgements to:
The authors gratefully also acknowledge the Research Council of Finland grant numbers 
339327, %``Carrington''
%1335554, %``ICT-SUNVAC'' already earlier
%1347795, % ``HISSA''
%1345701, % ``DAISY''
309937, %TEMPO
%322544, %Lucile
%328893, %Vertti
%338629, %AERGELC'H
3189131, %PROFI4
and 339756). %``KIMCHI''
The Finnish Centre of Excellence in Research of Sustainable Space, funded through the Academy of Finland grant 352846, supports Vlasiator development and science as well.

The authors would like to thank the LUMI Porting Program 2023 (\url{https://lumi-supercomputer.eu/porting-program-2023-5/}, with support from both LUMI system administrators and AMD / Advanced Micro Devices, Inc
, with particular thans to Samuel Antao at AMD for assisting in compiling, running, and managing memory on the LUMI-G supercomputer. 
Several GPU porting hackathons organized by CSC -- IT Center for Science Ltd have also been instrumental in guiding Vlasiator's GPU development. The simulations for this publication were run on the ``Mahti'' and ``LUMI'' Machines at the CSC site in Kajaani, Finland. %Porting to LUMI and AMD hardware was supported by CSC Hackathons as well as the LUMI porting program together with AMD Inc.

%\bibliographystyle{abbrvnat}
%\bibliographystyle{model1-num-names}
%\nocite{*}

\def\myitemsep{5pt}
%\bibliographystyle{aasjournal}
%\bibliography{solar_swqu}
\providecommand{\newblock}{}
%\begin{thebibliography}{10}
%\expandafter\ifx\csname url\endcsname\relax
%  \def\url#1{{\tt #1}}\fi
%\expandafter\ifx\csname urlprefix\endcsname\relax\def\urlprefix{URL }\fi
%\providecommand{\eprint}[2][]{\url{#2}}
% Bibliography created with iopart-num v2.1
% /biblio/bibtex/contrib/iopart-num

\bibliography{GPUpaper.bib}

\begin{thebibliography}{49}
\providecommand{\natexlab}[1]{#1}
\providecommand{\url}[1]{\texttt{#1}}
\expandafter\ifx\csname urlstyle\endcsname\relax
  \providecommand{\doi}[1]{doi: #1}\else
  \providecommand{\doi}{doi: \begingroup \urlstyle{rm}\Url}\fi

\bibitem[Alpay et~al.(2022)Alpay, Soproni, W{\"u}nsche, and Heuveline]{alpay2022exploring}
Aksel Alpay, B{\'a}lint Soproni, Holger W{\"u}nsche, and Vincent Heuveline.
\newblock Exploring the possibility of a hipsycl-based implementation of oneapi.
\newblock In \emph{International Workshop on OpenCL}, pages 1--12, 2022.

\bibitem[Asaduzzaman et~al.(2021)Asaduzzaman, Trent, Osborne, Aldershof, and Sibai]{asaduzzaman2021impact}
Abu Asaduzzaman, Alec Trent, S~Osborne, C~Aldershof, and Fadi~N Sibai.
\newblock Impact of cuda and opencl on parallel and distributed computing.
\newblock In \emph{2021 8th International Conference on Electrical and Electronics Engineering (ICEEE)}, pages 238--242. IEEE, 2021.

\bibitem[Battarbee et~al.(2020)Battarbee, Blanco-Cano, Turc, Kajdi\v{c}, Johlander, Tarvus, Fuselier, Trattner, Alho, Brito, Ganse, Pfau-Kempf, Akhavan-Tafti, Karlsson, Raptis, Dubart, Grandin, Suni, and Palmroth]{Battarbee_2020_Helium}
Markus Battarbee, Xochitl Blanco-Cano, Lucile Turc, P.~Kajdi\v{c}, Andreas Johlander, Vertti Tarvus, Stephen Fuselier, Karlheinz Trattner, Markku Alho, Thiago Brito, Urs Ganse, Yann Pfau-Kempf, Mojtaba Akhavan-Tafti, T.~Karlsson, Savvas Raptis, Maxime Dubart, Maxime Grandin, Jonas Suni, and Minna Palmroth.
\newblock Helium in the earth's foreshock: a global vlasiator survey.
\newblock \emph{Annales Geophysicae}, 38\penalty0 (5):\penalty0 1081--1099, 2020.
\newblock \doi{10.5194/angeo-38-1081-2020}.

\bibitem[Battarbee et~al.(2021)Battarbee, Brito, Alho, Pfau-Kempf, Grandin, Ganse, Papadakis, Johlander, Turc, Dubart, and Palmroth]{Battarbee_2021_evlasiator}
Markus Battarbee, Thiago Brito, Markku Alho, Yann Pfau-Kempf, Maxime Grandin, Urs Ganse, Konstantinos Papadakis, Andreas Johlander, Lucile Turc, Maxime Dubart, and Minna Palmroth.
\newblock Vlasov simulation of electrons in the context of hybrid global models: an {eVlasiator} approach.
\newblock \emph{Annales Geophysicae}, 39\penalty0 (1):\penalty0 85--103, January 2021.
\newblock \doi{10.5194/angeo-39-85-2021}.

\bibitem[Battarbee et~al.(2024)Battarbee, Papadakis, Ganse, Hokkanen, Antao, Kotipalo, Pfau-Kempf, and Alho]{vlasiatorgpucode}
Markus Battarbee, Kostis Papadakis, Urs Ganse, Jaro Hokkanen, Samuel Antao, Leo Kotipalo, Yann Pfau-Kempf, and Markku Alho.
\newblock {fmihpc/vlasiator: Vlasiator GPU pre-release v1}, 6 2024.
\newblock URL \url{https://doi.org/10.5281/zenodo.11472029}.

\bibitem[Beckingsale et~al.(2019)Beckingsale, Burmark, Hornung, Jones, Killian, Kunen, Pearce, Robinson, Ryujin, and Scogland]{RAJA}
D.~A. Beckingsale, J.~Burmark, R.~Hornung, H.~Jones, W.~Killian, A.~J. Kunen, O.~Pearce, P.~Robinson, B.~S. Ryujin, and T.~R.~W. Scogland.
\newblock {RAJA}: Portable performance for large-scale scientific applications.
\newblock In \emph{IEEE/ACM International Workshop on Performance, Portability and Productivity in HPC (P3HPC)}, volume 4,37, page 1370. The Open Journal, May 2019.
\newblock \doi{10.1109/P3HPC49587.2019.00012}.
\newblock URL \url{https://conferences.computer.org/sc19w/2019/#!/toc/14}.

\bibitem[Beckingsale et~al.(2020)Beckingsale, McFadden, Dahm, Pankajakshan, and Hornung]{umpire}
D.~A. Beckingsale, M.~J. McFadden, J.~P.~S. Dahm, R.~Pankajakshan, and R.~D. Hornung.
\newblock Umpire: Application-focused management and coordination of complex hierarchical memory.
\newblock \emph{IBM Journal of Research and Development}, 64\penalty0 (3/4):\penalty0 00:1--00:10, 2020.
\newblock \doi{10.1147/JRD.2019.2954403}.

\bibitem[Boman et~al.(2012)Boman, Catalyurek, Chevalier, and Devine]{Boman_2012_Zoltan}
E.~G. Boman, U.~V. Catalyurek, C.~Chevalier, and K.~D. Devine.
\newblock The {Z}oltan and {I}sorropia parallel toolkits for combinatorial scientific computing: Partitioning, ordering, and coloring.
\newblock \emph{Scientific Programming}, 20\penalty0 (2):\penalty0 129--150, 2012.
\newblock \doi{10.3233/SPR-2012-0342}.

\bibitem[Childs et~al.(2012)Childs, Brugger, Whitlock, Meredith, Ahern, Pugmire, Biagas, Miller, Harrison, Weber, Krishnan, Fogal, Sanderson, Garth, Bethel, Camp, R\"{u}bel, Durant, Favre, and Navr\'{a}til]{HPV:VisIt}
Hank Childs, Eric Brugger, Brad Whitlock, Jeremy Meredith, Sean Ahern, David Pugmire, Kathleen Biagas, Mark Miller, Cyrus Harrison, Gunther~H. Weber, Hari Krishnan, Thomas Fogal, Allen Sanderson, Christoph Garth, E.~Wes Bethel, David Camp, Oliver R\"{u}bel, Marc Durant, Jean~M. Favre, and Paul Navr\'{a}til.
\newblock Visit: An end-user tool for visualizing and analyzing very large data.
\newblock In \emph{High Performance Visualization--Enabling Extreme-Scale Scientific Insight}, pages 357--372. {}, October 2012.
\newblock \doi{10.1201/b12985}.

\bibitem[Christgau and Steinke(2020)]{christgau2020porting}
Steffen Christgau and Thomas Steinke.
\newblock Porting a legacy cuda stencil code to oneapi.
\newblock In \emph{2020 IEEE International Parallel and Distributed Processing Symposium Workshops (IPDPSW)}, pages 359--367. IEEE, 2020.

\bibitem[Costanzo et~al.(2021)Costanzo, Rucci, Sanchez, and Naiouf]{costanzo2021early}
Manuel Costanzo, Enzo Rucci, Carlos~Garc{\'\i}a Sanchez, and Marcelo Naiouf.
\newblock Early experiences migrating cuda codes to oneapi.
\newblock \emph{arXiv preprint arXiv:2105.13489}, 2021.

\bibitem[Einkemmer and Ostermann(2014)]{Einkemmer_2014_StrangSplitting}
Lukas Einkemmer and Alexander Ostermann.
\newblock Convergence analysis of strang splitting for vlasov-type equations.
\newblock \emph{{SIAM} Journal on Numerical Analysis}, 52\penalty0 (1):\penalty0 140--155, January 2014.
\newblock \doi{10.1137/130918599}.
\newblock URL \url{https://doi.org/10.1137\%2F130918599}.

\bibitem[{Evans}(2006)]{jemalloc}
J.~{Evans}.
\newblock A scalable concurrent {malloc(3)} implementation for {FreeBSD}.
\newblock In \emph{proceedings of {BSDCan} - The Technical {BSD} Conference}, April 2006.
\newblock URL \url{https://people.freebsd.org/~jasone/jemalloc/bsdcan2006/jemalloc.pdf}.
\newblock Referenced on May 22nd 2024.

\bibitem[Frank and Reich(2004)]{FrankReich2004}
J.~Frank and S.~Reich.
\newblock \emph{On Spurious Reflections, Nonuniform Grids and Finite Difference Discretizations of Wave Equations}.
\newblock CWI, Amsterdam, The Netherlands, 2004.

\bibitem[Ganse et~al.(2023)Ganse, Koskela, Battarbee, Pfau-Kempf, Papadakis, Alho, Bussov, Cozzani, Dubart, George, Gordeev, Grandin, Horaites, Suni, Tarvus, Kebede, Turc, Zhou, and Palmroth]{Ganse_2023_POP6D}
Urs Ganse, Tuomas Koskela, Markus Battarbee, Yann Pfau-Kempf, Konstantinos Papadakis, Markku Alho, Maarja Bussov, Giulia Cozzani, Maxime Dubart, Harriet George, Evgeny Gordeev, Maxime Grandin, Konstantinos Horaites, Jonas Suni, Vertti Tarvus, Fasil~Tesema Kebede, Lucile Turc, Hongyang Zhou, and Minna Palmroth.
\newblock {Enabling technology for global 3D + 3V hybrid-Vlasov simulations of near-Earth space}.
\newblock \emph{Physics of Plasmas}, 30\penalty0 (4):\penalty0 042902, 04 2023.
\newblock ISSN 1070-664X.
\newblock \doi{10.1063/5.0134387}.
\newblock URL \url{https://doi.org/10.1063/5.0134387}.

\bibitem[Ganse et~al.(2024)Ganse, Pfau-Kempf, Zhou, Juusola, Workayehu, Kebede, Papadakis, Grandin, Alho, Battarbee, Dubart, Koikkalainen, Kotipalo, Lalag\"ue, Suni, Tarvus, Horaites, and Palmroth]{ganse2024ionosphere}
Urs Ganse, Yann Pfau-Kempf, Hongyang Zhou, Liisa Juusola, Abiyot Workayehu, Fasil Kebede, Konstantinos Papadakis, Maxime Grandin, Markku Alho, Markus Battarbee, Maxime Dubart, Venla Koikkalainen, Leo Kotipalo, Arnaud Lalag\"ue, Jonas Suni, Vertti Tarvus, Konstantinos Horaites, and Minna Palmroth.
\newblock The vlasiator ionosphere -- coupling a magnetospheric hybrid-vlasov simulation with a height-integrated ionosphere model.
\newblock \emph{submitted to Annales Geophysicae}, 2024.

\bibitem[Grandin et~al.(2023)Grandin, Luttikhuis, Battarbee, Cozzani, Zhou, Turc, Pfau-Kempf, George, Horaites, Gordeev, Ganse, Papadakis, Alho, Tesema, Suni, Dubart, Tarvus, and Palmroth]{Grandin2023}
Maxime Grandin, Thijs Luttikhuis, Markus Battarbee, Giulia Cozzani, Hongyang Zhou, Lucile Turc, Yann Pfau-Kempf, Harriet George, Konstantinos Horaites, Evgeny Gordeev, Urs Ganse, Konstantinos Papadakis, Markku Alho, Fasil Tesema, Jonas Suni, Maxime Dubart, Vertti Tarvus, and Minna Palmroth.
\newblock {First 3D hybrid-Vlasov global simulation of auroral proton precipitation and comparison with satellite observations}.
\newblock \emph{Journal of Space Weather and Space Climate}, 13:\penalty0 20, 2023.
\newblock ISSN 2115-7251.
\newblock \doi{10.1051/SWSC/2023017}.

\bibitem[Honkonen et~al.(2013)Honkonen, {von Alfthan}, Sandroos, Janhunen, and Palmroth]{Honkonen_2013_DCCRG}
Ilja Honkonen, Sebastian {von Alfthan}, Arto Sandroos, Pekka Janhunen, and Minna Palmroth.
\newblock Parallel grid library for rapid and flexible simulation development.
\newblock \emph{Computer Physics Communications}, 184\penalty0 (4):\penalty0 1297 -- 1309, 2013.
\newblock ISSN 0010-4655.
\newblock \doi{10.1016/j.cpc.2012.12.017}.

\bibitem[Kotipalo et~al.(2024)Kotipalo, Battarbee, Pfau-Kempf, and Palmroth]{Kotipalo_2024_dAMR}
L.~Kotipalo, M.~Battarbee, Y.~Pfau-Kempf, and M.~Palmroth.
\newblock Adaptive mesh refinement in vlasiator.
\newblock \emph{Submitted to Jour. Nal.}, 2024.

\bibitem[Londrillo and Del~Zanna(2004)]{Londrillo_2004_fieldsolver}
P~Londrillo and Luca Del~Zanna.
\newblock On the divergence-free condition in godunov-type schemes for ideal magnetohydrodynamics: the upwind constrained transport method.
\newblock \emph{Journal of Computational Physics}, 195\penalty0 (1):\penalty0 17--48, March 2004.
\newblock \doi{10.1016/j.jcp.2003.09.016}.
\newblock URL \url{http://www.sciencedirect.com/science/article/pii/S0021999103005102}.

\bibitem[Luebke(2008)]{luebke2008cuda}
David Luebke.
\newblock Cuda: Scalable parallel programming for high-performance scientific computing.
\newblock In \emph{2008 5th IEEE international symposium on biomedical imaging: from nano to macro}, pages 836--838. IEEE, 2008.

\bibitem[{Matthes} et~al.(2017){Matthes}, {Widera}, {Zenker}, {Worpitz}, {Huebl}, and {Bussmann}]{alpaka_MathesP3MA2017}
A.~{Matthes}, R.~{Widera}, E.~{Zenker}, B.~{Worpitz}, A.~{Huebl}, and M.~{Bussmann}.
\newblock Tuning and optimization for a variety of many-core architectures without changing a single line of implementation code using the alpaka library.
\newblock In \emph{ISC High Performance Workshops 2017, LNCS 10524, pp. 496-514, 2017}, Jun 2017.
\newblock URL \url{https://arxiv.org/abs/1706.10086}.

\bibitem[Nishikawa et~al.(2021)Nishikawa, Du{\c t}an, K{\"o}hn, and Mizuno]{nishikawa_pic_2021}
Kenichi Nishikawa, Ioana Du{\c t}an, Christoph K{\"o}hn, and Yosuke Mizuno.
\newblock {{PIC}} methods in astrophysics: Simulations of relativistic jets and kinetic physics in astrophysical systems.
\newblock \emph{Living Reviews in Computational Astrophysics}, 7\penalty0 (1):\penalty0 1, July 2021.
\newblock ISSN 2365-0524.
\newblock \doi{10.1007/s41115-021-00012-0}.

\bibitem[Palmroth et~al.(2018)Palmroth, Ganse, Pfau-Kempf, Battarbee, Turc, Brito, Grandin, Hoilijoki, Sandroos, and von Alfthan]{Palmroth_2018_LRCA}
Minna Palmroth, Urs Ganse, Yann Pfau-Kempf, Markus Battarbee, Lucile Turc, Thiago Brito, Maxime Grandin, Sanni Hoilijoki, Arto Sandroos, and Sebastian von Alfthan.
\newblock Vlasov methods in space physics and astrophysics.
\newblock \emph{Living Reviews in Computational Astrophysics}, 4\penalty0 (1):\penalty0 1, August 2018.
\newblock ISSN 2365-0524.
\newblock \doi{10.1007/s41115-018-0003-2}.

\bibitem[Palmroth et~al.(2023)Palmroth, Pulkkinen, Ganse, Pfau-Kempf, Koskela, Zaitsev, Alho, Cozzani, Turc, Battarbee, Dubart, George, Gordeev, Grandin, Horaites, Osmane, Papadakis, Suni, Tarvus, Zhou, and Nakamura]{Palmroth2023}
Minna Palmroth, Tuija~I. Pulkkinen, Urs Ganse, Yann Pfau-Kempf, Tuomas Koskela, Ivan Zaitsev, Markku Alho, Giulia Cozzani, Lucile Turc, Markus Battarbee, Maxime Dubart, Harriet George, Evgeniy Gordeev, Maxime Grandin, Konstantinos Horaites, Adnane Osmane, Konstantinos Papadakis, Jonas Suni, Vertti Tarvus, Hongyang Zhou, and Rumi Nakamura.
\newblock {Magnetotail plasma eruptions driven by magnetic reconnection and kinetic instabilities}.
\newblock \emph{Nature Geoscience 2023 16:7}, 16\penalty0 (7):\penalty0 570--576, 6 2023.
\newblock ISSN 1752-0908.
\newblock \doi{10.1038/s41561-023-01206-2}.

\bibitem[Papadakis et~al.(2022)Papadakis, Pfau-Kempf, Ganse, Battarbee, Alho, Grandin, Dubart, Turc, Zhou, Horaites, Zaitsev, Cozzani, Bussov, Gordeev, Tesema, George, Suni, Tarvus, and Palmroth]{Papadakis_2022}
K.~Papadakis, Y.~Pfau-Kempf, U.~Ganse, M.~Battarbee, M.~Alho, M.~Grandin, M.~Dubart, L.~Turc, H.~Zhou, K.~Horaites, I.~Zaitsev, G.~Cozzani, M.~Bussov, E.~Gordeev, F.~Tesema, H.~George, J.~Suni, V.~Tarvus, and M.~Palmroth.
\newblock Spatial filtering in a 6d hybrid-vlasov scheme for alleviating amr artifacts: a case study with vlasiator, versions 5.0, 5.1, 5.2.1.
\newblock \emph{Geoscientific Instrumntation, Methods and Data Systems}, 2022:\penalty0 1--18, 2022.
\newblock \doi{10.5194/gmd-15-7903-2022}.

\bibitem[Papadakis et~al.(2024{\natexlab{a}})Papadakis, Battarbee, Ganse, Pfau-Kempf, and Palmroth]{Papadakis_2024_Hashinator}
K.~Papadakis, M.~Battarbee, U.~Ganse, Y.~Pfau-Kempf, and M.~Palmroth.
\newblock Hashinator: A portable hybrid hashmap designed for heterogeneous high performance computing.
\newblock \emph{Accepted for publication in Frontiers in Computer Science Software}, 2024{\natexlab{a}}.
\newblock \doi{10.3389/fcomp.2024.1407365}.

\bibitem[Papadakis et~al.(2024{\natexlab{b}})Papadakis, Battarbee, and Widera]{HashinatorZenodo}
Kostis Papadakis, Markus Battarbee, and René Widera.
\newblock fmihpc/hashinator: v1.0.1 hashinator stable, 2024{\natexlab{b}}.
\newblock URL \url{https://zenodo.org/doi/10.5281/zenodo.11396297}.

\bibitem[Pekkilä et~al.(2022)Pekkilä, Väisälä, Käpylä, Rheinhardt, and Lappi]{PEKKILA2022102904}
Johannes Pekkilä, Miikka~S. Väisälä, Maarit~J. Käpylä, Matthias Rheinhardt, and Oskar Lappi.
\newblock Scalable communication for high-order stencil computations using cuda-aware mpi.
\newblock \emph{Parallel Computing}, 111:\penalty0 102904, 2022.
\newblock ISSN 0167-8191.
\newblock \doi{https://doi.org/10.1016/j.parco.2022.102904}.
\newblock URL \url{https://www.sciencedirect.com/science/article/pii/S0167819122000102}.

\bibitem[Pfau-Kempf et~al.(2018)Pfau-Kempf, Battarbee, Ganse, Hoilijoki, Turc, von Alfthan, Vainio, and Palmroth]{PfauKempf_2018_scaling}
Yann Pfau-Kempf, Markus Battarbee, Urs Ganse, Sanni Hoilijoki, Lucile Turc, Sebastian von Alfthan, Rami Vainio, and Minna Palmroth.
\newblock On the importance of spatial and velocity resolution in the hybrid-vlasov modeling of collisionless shocks.
\newblock \emph{Frontiers in Physics}, 2018.
\newblock \doi{10.3389/fphy.2018.00044}.

\bibitem[Pfau-Kempf et~al.(2020)Pfau-Kempf, Palmroth, Johlander, Turc, Alho, Battarbee, Dubart, Grandin, and Ganse]{PfauKempf_2020_2.9D}
Yann Pfau-Kempf, Minna Palmroth, Andreas Johlander, Lucile Turc, Markku Alho, Markus Battarbee, Maxime Dubart, Maxime Grandin, and Urs Ganse.
\newblock Hybrid-vlasov modeling of three-dimensional dayside magnetopause reconnection.
\newblock \emph{Physics of Plasmas}, 27\penalty0 (9):\penalty0 092903, 2020.
\newblock \doi{10.1063/5.0020685}.

\bibitem[Pfau-Kempf et~al.(2022)Pfau-Kempf, von Alfthan, Ganse, Sandroos, Battarbee, Koskela, Hannuksela, Honkonen, Papadakis, Kotipalo, Zhou, Grandin, Pokhotelov, and Alho]{vlasiatorcode}
Yann Pfau-Kempf, Sebastian von Alfthan, Urs Ganse, Arto Sandroos, Markus Battarbee, Tuomas Koskela, Otto~Akseli Hannuksela, Ilja Honkonen, Kostis Papadakis, Leo Kotipalo, Hongyang Zhou, Maxime Grandin, Dimitry Pokhotelov, and Markku Alho.
\newblock {fmihpc/vlasiator: Vlasiator}, 6 2022.
\newblock URL \url{https://doi.org/10.5281/zenodo.3640593}.

\bibitem[Sandroos et~al.(2013)Sandroos, Honkonen, {von Alfthan}, and Palmroth]{SANDROOS2013306}
A.~Sandroos, I.~Honkonen, S.~{von Alfthan}, and M.~Palmroth.
\newblock Multi-gpu simulations of vlasov’s equation using vlasiator.
\newblock \emph{Parallel Computing}, 39\penalty0 (8):\penalty0 306--318, 2013.
\newblock ISSN 0167-8191.
\newblock \doi{https://doi.org/10.1016/j.parco.2013.05.001}.
\newblock URL \url{https://www.sciencedirect.com/science/article/pii/S0167819113000574}.

\bibitem[Strang(1968)]{Strang_1968_Splitting}
Gilbert Strang.
\newblock On the construction and comparison of difference schemes.
\newblock \emph{{SIAM} Journal on Numerical Analysis}, 5\penalty0 (3):\penalty0 506--517, September 1968.
\newblock \doi{10.1137/0705041}.
\newblock URL \url{https://doi.org/10.1137\%2F0705041}.

\bibitem[Tarvus et~al.(2024)Tarvus, Turc, Zhou, Nakamura, Settino, Blasl, Cozzani, Ganse, Pfau-Kempf, Alho, Battarbee, Bussov, Dubart, Gordeev, Kebede, Papadakis, Suni, Zaitsev, and Palmroth]{Tarvus2024}
V.~Tarvus, L.~Turc, H.~Zhou, T.~Nakamura, A.~Settino, K.~Blasl, G.~Cozzani, U.~Ganse, Y.~Pfau-Kempf, M.~Alho, M.~Battarbee, M.~Bussov, M.~Dubart, E.~Gordeev, F.~Kebede, K.~Papadakis, J.~Suni, I.~Zaitsev, and M.~Palmroth.
\newblock Hybrid-vlasov modelling of ion velocity distribution functions associated with the kelvin-helmholtz instability with a density and temperature asymmetry.
\newblock \emph{Under revision at the astrophysical journal}, 2024.

\bibitem[Trott et~al.(2022)Trott, Lebrun-Grandié, Arndt, Ciesko, Dang, Ellingwood, Gayatri, Harvey, Hollman, Ibanez, Liber, Madsen, Miles, Poliakoff, Powell, Rajamanickam, Simberg, Sunderland, Turcksin, and Wilke]{kokkos9485033}
Christian~R. Trott, Damien Lebrun-Grandié, Daniel Arndt, Jan Ciesko, Vinh Dang, Nathan Ellingwood, Rahulkumar Gayatri, Evan Harvey, Daisy~S. Hollman, Dan Ibanez, Nevin Liber, Jonathan Madsen, Jeff Miles, David Poliakoff, Amy Powell, Sivasankaran Rajamanickam, Mikael Simberg, Dan Sunderland, Bruno Turcksin, and Jeremiah Wilke.
\newblock Kokkos 3: Programming model extensions for the exascale era.
\newblock \emph{IEEE Transactions on Parallel and Distributed Systems}, 33\penalty0 (4):\penalty0 805--817, 2022.
\newblock \doi{10.1109/TPDS.2021.3097283}.

\bibitem[Tsai et~al.(2021)Tsai, Cojean, Ribizel, and Anzt]{tsai2021}
Yuhsiang~M. Tsai, Terry Cojean, Tobias Ribizel, and Hartwig Anzt.
\newblock Preparing ginkgo for amd gpus -- a testimonial on porting cuda code to hip.
\newblock In Bartosz Balis, Dora B.~Heras, Laura Antonelli, Andrea Bracciali, Thomas Gruber, Jin Hyun-Wook, Michael Kuhn, Stephen~L. Scott, Didem Unat, and Roman Wyrzykowski, editors, \emph{Euro-Par 2020: Parallel Processing Workshops}, pages 109--121, Cham, 2021. Springer International Publishing.
\newblock ISBN 978-3-030-71593-9.

\bibitem[Videau(2022)]{HIP-oneapi}
Brice Videau.
\newblock {Bringing HIP to oneAPI}, 2022.
\newblock Accessed on December 1. 2023.

\bibitem[Vijayaraghavan et~al.(2017)Vijayaraghavan, Eckert, Loh, Schulte, Ignatowski, Beckmann, Brantley, Greathouse, Huang, Karunanithi, Kayiran, Meswani, Paul, Poremba, Raasch, Reinhardt, Sadowski, and Sridharan]{7920816}
Thiruvengadam Vijayaraghavan, Yasuko Eckert, Gabriel~H. Loh, Michael~J. Schulte, Mike Ignatowski, Bradford~M. Beckmann, William~C. Brantley, Joseph~L. Greathouse, Wei Huang, Arun Karunanithi, Onur Kayiran, Mitesh Meswani, Indrani Paul, Matthew Poremba, Steven Raasch, Steven~K. Reinhardt, Greg Sadowski, and Vilas Sridharan.
\newblock Design and analysis of an apu for exascale computing.
\newblock In \emph{2017 IEEE International Symposium on High Performance Computer Architecture (HPCA)}, pages 85--96, 2017.
\newblock \doi{10.1109/HPCA.2017.42}.

\bibitem[Vlasov(1961)]{Vlasov_1961_Vlasov}
Anatoly~A. Vlasov.
\newblock \emph{Many-particle Theory and Its Application to Plasma}.
\newblock Gordon \& Breach Science Publishers Ltd, 1961.
\newblock ISBN 9780677203300.

\bibitem[von Alfthan et~al.(2014)von Alfthan, Pokhotelov, Kempf, Hoilijoki, Honkonen, Sandroos, and Palmroth]{Alfthan_2014_vlasiator}
S.~von Alfthan, D.~Pokhotelov, Y.~Kempf, S.~Hoilijoki, I.~Honkonen, A.~Sandroos, and M.~Palmroth.
\newblock Vlasiator: First global hybrid-vlasov simulations of earth's foreshock and magnetosheath.
\newblock \emph{Journal of Atmospheric and Solar-Terrestrial Physics}, 120\penalty0 (0):\penalty0 24 -- 35, 2014.
\newblock ISSN 1364-6826.
\newblock \doi{10.1016/j.jastp.2014.08.012}.

\bibitem[von Alfthan(2023)]{phiprof}
Sebastian von Alfthan.
\newblock Phiprof -- parallel hierarchical profiler.
\newblock \url{https://github.com/fmihpc/phiprof}, 2023.

\bibitem[Walker and Dongarra(1996)]{walker1996mpi}
David~W Walker and Jack~J Dongarra.
\newblock Mpi: a standard message passing interface.
\newblock \emph{Supercomputer}, 12:\penalty0 56--68, 1996.

\bibitem[Wang et~al.(2014)Wang, Ding, Lee, Kato, and Zhang]{10.1145/2591971.2592002}
Kaibo Wang, Xiaoning Ding, Rubao Lee, Shinpei Kato, and Xiaodong Zhang.
\newblock Gdm: device memory management for gpgpu computing.
\newblock In \emph{The 2014 ACM International Conference on Measurement and Modeling of Computer Systems}, SIGMETRICS '14, page 533–545, New York, NY, USA, 2014. Association for Computing Machinery.
\newblock ISBN 9781450327893.
\newblock \doi{10.1145/2591971.2592002}.
\newblock URL \url{https://doi.org/10.1145/2591971.2592002}.

\bibitem[Yee(1966)]{yee_numerical_1966}
Kane Yee.
\newblock Numerical solution of initial boundary value problems involving maxwell's equations in isotropic media.
\newblock \emph{IEEE Transactions on Antennas and Propagation}, 14\penalty0 (3):\penalty0 302--307, May 1966.
\newblock ISSN 1558-2221.
\newblock \doi{10.1109/TAP.1966.1138693}.

\bibitem[Zerroukat and Allen(2012)]{Zerroukat_2012_SLICE3D}
M.~Zerroukat and T.~Allen.
\newblock A three-dimensional monotone and conservative semi-lagrangian scheme (slice-3d) for transport problems.
\newblock \emph{Quarterly Journal of the Royal Meteorological Society}, 138\penalty0 (667):\penalty0 1640--1651, 2012.
\newblock \doi{10.1002/qj.1902}.

\bibitem[Zhang et~al.(2019)Zhang, Almgren, Beckner, Bell, Blaschke, Chan, Day, Friesen, Gott, Graves, Katz, Myers, Nguyen, Nonaka, Rosso, Williams, and Zingale]{AMReX_JOSS}
Weiqun Zhang, Ann Almgren, Vince Beckner, John Bell, Johannes Blaschke, Cy~Chan, Marcus Day, Brian Friesen, Kevin Gott, Daniel Graves, Max Katz, Andrew Myers, Tan Nguyen, Andrew Nonaka, Michele Rosso, Samuel Williams, and Michael Zingale.
\newblock {AMReX}: a framework for block-structured adaptive mesh refinement.
\newblock \emph{Journal of Open Source Software}, 4\penalty0 (37):\penalty0 1370, May 2019.
\newblock \doi{10.21105/joss.01370}.
\newblock URL \url{https://doi.org/10.21105/joss.01370}.

\bibitem[Zheng et~al.(2016)Zheng, Nellans, Zulfiqar, Stephenson, and Keckler]{7446077}
Tianhao Zheng, David Nellans, Arslan Zulfiqar, Mark Stephenson, and Stephen~W. Keckler.
\newblock Towards high performance paged memory for gpus.
\newblock In \emph{2016 IEEE International Symposium on High Performance Computer Architecture (HPCA)}, pages 345--357, 2016.
\newblock \doi{10.1109/HPCA.2016.7446077}.

\bibitem[Zhu et~al.(2018)Zhu, Zhuo, Wang, Chen, and Xie]{8293707}
Maohua Zhu, Youwei Zhuo, Chao Wang, Wenguang Chen, and Yuan Xie.
\newblock Performance evaluation and optimization of hbm-enabled gpu for data-intensive applications.
\newblock \emph{IEEE Transactions on Very Large Scale Integration (VLSI) Systems}, 26\penalty0 (5):\penalty0 831--840, 2018.
\newblock \doi{10.1109/TVLSI.2018.2791442}.

\end{thebibliography}
\end{document}